%% file: ms.tex
\documentclass[%
 reprint,
 amsmath,amssymb,
 aps,
]{revtex4-2}

\usepackage{graphicx}
\usepackage{dcolumn}
\usepackage[hidelinks]{hyperref}
\usepackage{bm}
\usepackage{xcolor}
\usepackage{color,soul}

\begin{document}

\preprint{APS/123-QED}

\include{text_main} 


\clearpage
\include{text_SI} 

\end{document}

%% file: text_main.tex
\title{Transition rates, survival probabilities, and quality of bias from time-dependent biased simulations}

\author{Karen Palacio-Rodriguez$^{1,2}$}
\author{Hadrien Vroylandt$^{3}$}
\author{Lukas S. Stelzl$^{4,5,6,7}$}
\author{Fabio Pietrucci$^{1}$}
\author{Gerhard Hummer$^{7,8}$}
\author{Pilar Cossio$^{2,7,9}$}
\email{pcossio@flatironinstitute.org}
\affiliation{$^{1}$Sorbonne Université, Institut de Minéralogie, de Physique des Matériaux et de Cosmochimie, IMPMC, F-75005 Paris, France}
\affiliation{$^{2}$Biophysics of Tropical Diseases Max Planck Tandem Group, University of Antioquia, Medellín, Colombia}
\affiliation{$^{3}$Sorbonne Université, Institut des sciences du calcul et des données, ISCD, F-75005 Paris, France}
\affiliation{$^{4}$Faculty of Biology, Johannes Gutenberg University Mainz, Gresemundweg 2, 55128 Mainz, Germany}
\affiliation{$^{5}$KOMET 1, Institute of Physics, Johannes Gutenberg University  Mainz, 55099 Mainz, Germany}
\affiliation{$^{6}$Institute of Molecular Biology (IMB), 55128 Mainz, Germany}
\affiliation{$^{7}$Department of Theoretical Biophysics, Max Planck Institute of Biophysics, Max-von-Laue Straße 3, 60438 Frankfurt am Main, Germany}
\affiliation{$^{8}$Institute for Biophysics, Goethe University Frankfurt, 60438 Frankfurt am Main, Germany}
\affiliation{$^{9}$Center for Computational Mathematics, Flatiron Institute, New York, USA}

\begin{abstract}
Simulations with an adaptive time-dependent bias, such as metadynamics, enable an efficient exploration of the conformational space of a system. However, the dynamic information of the system is altered by the bias. With infrequent metadynamics it is possible to recover the transition rate of crossing a barrier, if the collective variables are ideal and there is no bias deposition near the transition state. Unfortunately, for simulations of complex molecules, these conditions are not always fulfilled. To overcome these limitations, and  inspired by single-molecule force spectroscopy, we developed a method based on Kramers' theory for calculating the barrier-crossing rate when a time-dependent bias is added to the system. We assess the \textit{quality of the bias} parameter by measuring how efficiently the bias accelerates the transitions compared to ideal behavior. We present approximate analytical expressions of the survival probability that accurately reproduce the barrier-crossing time statistics, and enable the extraction of the unbiased transition rate even for challenging cases, where previous methods fail.
\end{abstract}

\maketitle

Kinetic rate constants are of fundamental importance by quantifying the speed of interconversion between metastable states in the description of physical phenomena. From protein folding to nucleation, the estimation of the transition rates allows us to understand the time scales of the events and the mechanistic implications. However, obtaining rate coefficients for rare events is not a trivial task. Computer-assisted methods have gained relevance in recent decades to predict kinetic properties \cite{Holenz2019,Rognan2017a,Talele2010,Jorgensen2004}. Among these methods, molecular dynamics (MD) simulations have been widely used to study the thermodynamic and kinetic behavior of molecular systems \cite{Bernetti2019,Hollingsworth2018,Ganesan2017}. Processes such as protein folding or ligand binding have been successfully studied by means of MD simulations \cite{Lindorff-Larsen2011,Piana2013,Chodera2014,Plattner2017,Tang2020a,Wolf2020}. However, MD has the limitation that the time scales of many rare events are not accessible by standard simulations, even using powerful supercomputers \cite{DeVivo2016b,Dickson2017,Ribeiro2019a}.

Enhanced sampling methods in combination with MD simulations have become useful alternatives for studying events that occur at long timescales \cite{Pietrucci2017}. These methods have been developed to accelerate the sampling of the conformational space, typically characterized by rugged landscapes and high energy barriers \cite{Bernardi2015b}. Among these methods, metadynamics \cite{Laio2002} (MetaD), is an enhanced-sampling technique where the conformational search is accelerated by adding a history-dependent bias potential to the force-field. The biasing potential is a function of collective variables (CVs) chosen to describe the degrees of freedom considered most relevant to the transition mechanism \cite{Laio2002,Aci-Seche2016}. MetaD has the advantage that, for a converged simulation and appropriate CVs \cite{Cavalli2015}, it is possible to directly recover the free energy profile of the system from the MetaD bias \cite{bussi2020}. However, a disadvantage of MetaD, as well as other enhanced sampling methods, is that information about the dynamics of the simulated system is corrupted, due to the sampling acceleration \cite{Cavalli2015}. Therefore, it may seem impossible to extract quantitative rate information from such simulations.

Nevertheless, several methods have been developed to estimate rate coefficients from enhanced-sampling simulations \cite{Dickson2017,camilloni2018,Kokh2018,Nunes2020}. Some involve the calculation of diffusion coefficients and the construction of Markov State Models \cite{Hummer2005,Marinelli2009,Tiwary2013,Stelzl2017a,Stelzl2017,Donati2018,Schafer2020b,Kieninger2020,Linker2020}. Among these methods infrequent metadynamics \cite{Tiwary2013} (iMetaD), has been widely used in recent years \cite{Tiwary2015a,Tiwary2017,Casasnovas2017,Sun2017a,Pramanik2019,Zou2020,LamimRibeiro2020,Shekhar2021}. iMetaD is based on the transition state theory, and employs an acceleration factor \cite{Grubmuller1995a,Voter1997a} extracted from MetaD simulations. The main idea is to deposit bias infrequently so that no bias is deposited in the region of the transition state. In this way the dynamics of the TS region is not corrupted \cite{Tiwary2013}, and it is possible to correct the escape times from a state. In addition to the slow biasing frequency, this approach also requires a small set of CVs that determine the relevant states and pathways of the system \cite{Dickson2017}. When these conditions are satisfied, the distribution of escape times follows a Poisson behavior \cite{Salvalaglio2014}. The reliability of the distribution of the rescaled escape times obtained with iMetaD is tested using the Kolmogorov-Smirnov test (KS), which compares the cumulative distribution function (CDF) of the rescaled (escape) times to that theoretically expected \cite{Salvalaglio2014}. Despite the usefulness of this, a major limitation is that it relies on ideal CVs, where the bias potential is zero at all the dividing surfaces \cite{Dickson2018,Khan2020}. Modifications of iMetaD have thus been proposed \cite{Callegari2017,Wang2018}. However, these methods do not directly compute the time-dependent rate or survival probably due to the bias.

Here, we take inspiration from dynamic force spectroscopy experiments, where a bias is ramped up with time similar to the simulations with a dynamic bias. For these experiments, accurate kinetic predictions of the force-dependent rates and transition probabilities \cite{Hummer2003,Dudko2006,Cossio2016} have been derived. By considering the biasing potential analogous to an external force, we introduce a physical model of barrier-crossing events in time-dependent biased simulations for computing directly the transition statistics. The major advantages are that one can extract the unbiased rate and, at the same time, assess the quality of the CVs in terms of their contribution to the bias acceleration. 

\begin{figure}[t!]
    \centering
    \includegraphics[width=8.6cm]{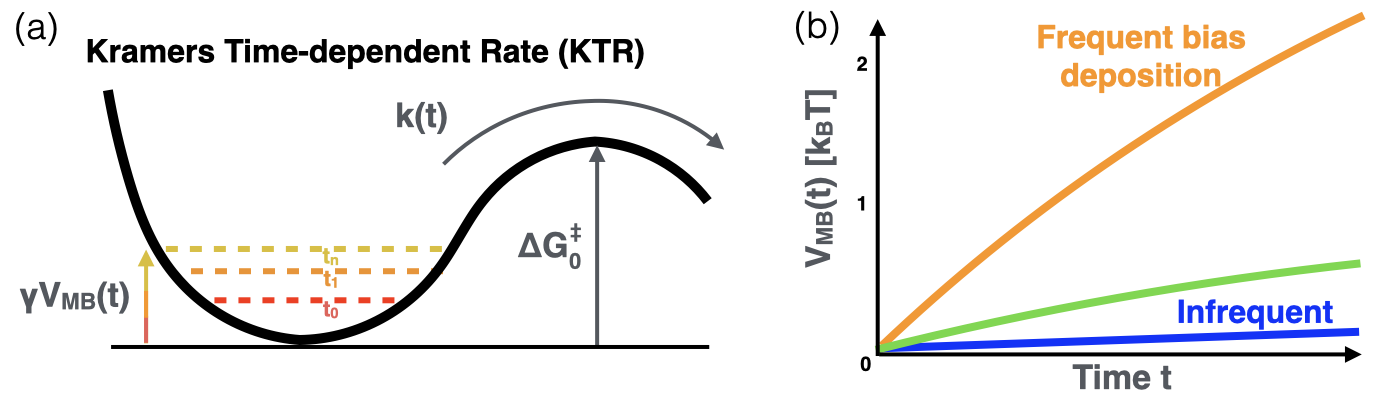}
    \caption{\textbf{Kramers time-dependent barrier-crossing rate for biased simulations.} (a) Schematic representation of the escape from a bottom-well with a time-dependent bias. $\gamma V_{\mathrm{MB}}(t)$ measures the effective contribution of the added bias-height toward lowering the effective barrier. (b) Examples of $V_{\mathrm{MB}}(t)$ for frequent and infrequent bias-deposition times. The results are, for the double-well potential example, $V_{\mathrm{MB}}(t)$ along $y$ with bias-deposition times $d_t=1$ (orange), $5$ (green), and $20$ (blue) ($\times 10^3$) steps averaged over multiple runs. }
    \label{fig1}
\end{figure}


For diffusive dynamics, and high barriers, Kramers' theory \cite{Kramers1940,Hanggi1990} is used to calculate the rate (\textit{i.e.} inverse of mean residence time) to cross a barrier along a coordinate
\begin{equation}
    k_0 = k_{\mathrm{pre}}e^{-\beta\Delta G^\ddagger_0}~,
    \label{eq:intrate}
\end{equation}
where $\Delta G^\ddagger_0$ is the barrier height (Fig. \ref{fig1}a), $\beta=1/k_BT$ is the inverse temperature, and $k_\mathrm{pre}$ is the pre-exponential factor that depends on the diffusion coefficient, shape of the bottom-well and barrier-top.

In MetaD short-range repulsive functions are deposited at regular time intervals within a low-dimensional CV space. Over time, the bias fills the well. This reduces the effective barrier experienced by the system which is thus crossed more rapidly. To describe this reduction of the barrier height, we use the time-dependent maximum bias (MB) averaged over multiple runs
\begin{equation}
 V_{\mathrm{MB}}(t) = \frac{1}{R} \sum_r \max_{t'\in[0,t]}\, V^r_B(t') ~,
\end{equation}
where $V^r_B(t)$ is the instantaneous bias at time $t$ for simulation run $r$, and $R$ is the total number of runs. $V_{\mathrm{MB}}(t)$ is the average maximum height of the biasing potential (\textit{i.e.}, the level of bias added to the bottom-well) up to time $t$. In the case of an ideal CV, $\Delta G^\ddagger_0 - V_{\mathrm{MB}}(t)$ would be the effective time-dependent barrier experienced by the biased system.

$V_{\mathrm{MB}}(t)$ depends on the shape of the potential surface, the bias-deposition time ($d_t$), and bias-deposition height, among others. In Fig. \ref{fig1}b, we present some examples of $V_{\mathrm{MB}}(t)$ for several $d_t$. The blue line shows the case for which iMetaD is valid. We will compute directly the statistics for barrier-crossing times from biased simulations, covering a wider range of $V_{\mathrm{MB}}(t)$, by explicitly taking into account their time dependence.

Assuming a quasi-adiabatic bias deposition, we apply Kramers' theory over the potential presented in Fig. \ref{fig1}a, to calculate the time-dependent rate of escape due to the bias acceleration
\begin{equation}
    k(t) = k_{\mathrm{pre}}e^{-\beta\Delta G^\ddagger_0+\beta\gamma V_{\mathrm{MB}}(t)} = k_0\,e^{\beta\gamma V_{\mathrm{MB}}(t)}~, 
    \label{eq:kt}
\end{equation}
where $k_0$ is the intrinsic rate (Eq. \ref{eq:intrate}), and we are assuming that $k_{\mathrm{pre}}$ does not change due to the bias. 

We introduce $\gamma~\in[0,1]$ as an additional parameter that measures how much of the bias contributes to the acceleration. For ideal CVs, \textit{i.e.}, where the added bias acts along the direction of the true transition and helps to lower the effective barrier, we expect $\gamma=1$. 
By contrast, we expect $\gamma=0$ for poorly chosen CVs, \textit{i.e.}, where the bias acts in directions orthogonal to the transition. We illustrate this behavior of $\gamma$ for a cusp-like harmonic double-well potential (shown in Supplementary Text Eq. 1) where parameter $a$ controls the separation between the wells along $x$, $i.e.,$ the quality of the $x$ CV. If $a\approx 0$, $x$ is a poor coordinate because the wells are not separated when projecting along this direction. 
In this case, if we bias along $x$, we show that $\gamma\approx 2a$ (and therefore $\gamma \rightarrow 0$). In contrast, for this potential, when biasing along the good CV $y$, $\gamma=1$ is derived.  Therefore, $\gamma$ is not simply an \textit{ad hoc} bias factor, but it measures a physical property that can be related to the quality of the CV.

Using Kramers' Time-dependent Rate (KTR), from Eq. \ref{eq:kt}, we calculate the survival probability by adapting the methods used for the analysis of single-molecule force spectroscopy experiments \cite{Hummer2003}, 
\begin{equation}
    S(t) = \exp\left(-\int_0^t k(t')dt'\right)=\exp\left(-k_0\int_0^t e^{\beta\gamma V_{\mathrm{MB}}(t')}dt'\right)\,.
    \label{eq:st}
\end{equation}
The survival probability depends on $V_{\mathrm{MB}}(t)$. For example, if the bias presents a logarithmic time dependence, $V_{\mathrm{MB}}(t)=a\log(1+b\,t)$ (\textit{e.g.}, orange line in Fig. \ref{fig1}b), then the survival probability is
\begin{equation}
    S(t) = \exp\left(\frac{k_0}{b(\beta\gamma a+1)}\left(1-(1+b\,t)^{\beta\gamma a+1}\right)\right)\,.
    \label{eq:stlog}
\end{equation}
In the Supplementary Text, the analytic expression for a linear time-dependent bias is presented. For more general $V_{\mathrm{MB}}(t)$, we can solve Eq. \ref{eq:st} numerically. 

Let us assume that there are $M+N$ independent biased simulations that start from the well bottom where $M$ cross the barrier and $N$ remain in the basin. By monitoring $V^r_B(t)$ over the runs, we can calculate $V_{\mathrm{MB}}(t)$ (Fig. \ref{fig1}b), and use it to calculate $S(t)$ (Eq. \ref{eq:st}). The transition rate $k_0$ and $\gamma$ can be extracted in two manners: \textit{i)} by calculating the cumulative distribution function (CDF) using the simulation barrier-crossing times (without rescaling) and fitting the CDF using $1-S(t)$, or \textit{ii)} by maximizing the likelihood function
\begin{equation}
    \mathcal{L}=\prod_{i \in \mathrm{events}}^M \frac{dS(t)}{dt}\Big|_{t=t_i} \prod_{j \in \mathrm{non-events}}^N S(t_j)~,
    \label{eq:L}
\end{equation}
where $i$ and $j$ account for events and non-events, respectively, $t_i$ is the escape time observed in the biased simulation $i$, and $t_j$ is the total simulation time for run $j$ that did not transition. To validate the statistics of the simulation barrier-crossing times, bootstrap analysis and KS-tests can be performed using the CDF (see the Supplementary Text).

\begin{figure}[htb!]
    \centering
    \includegraphics[width=8.6cm]{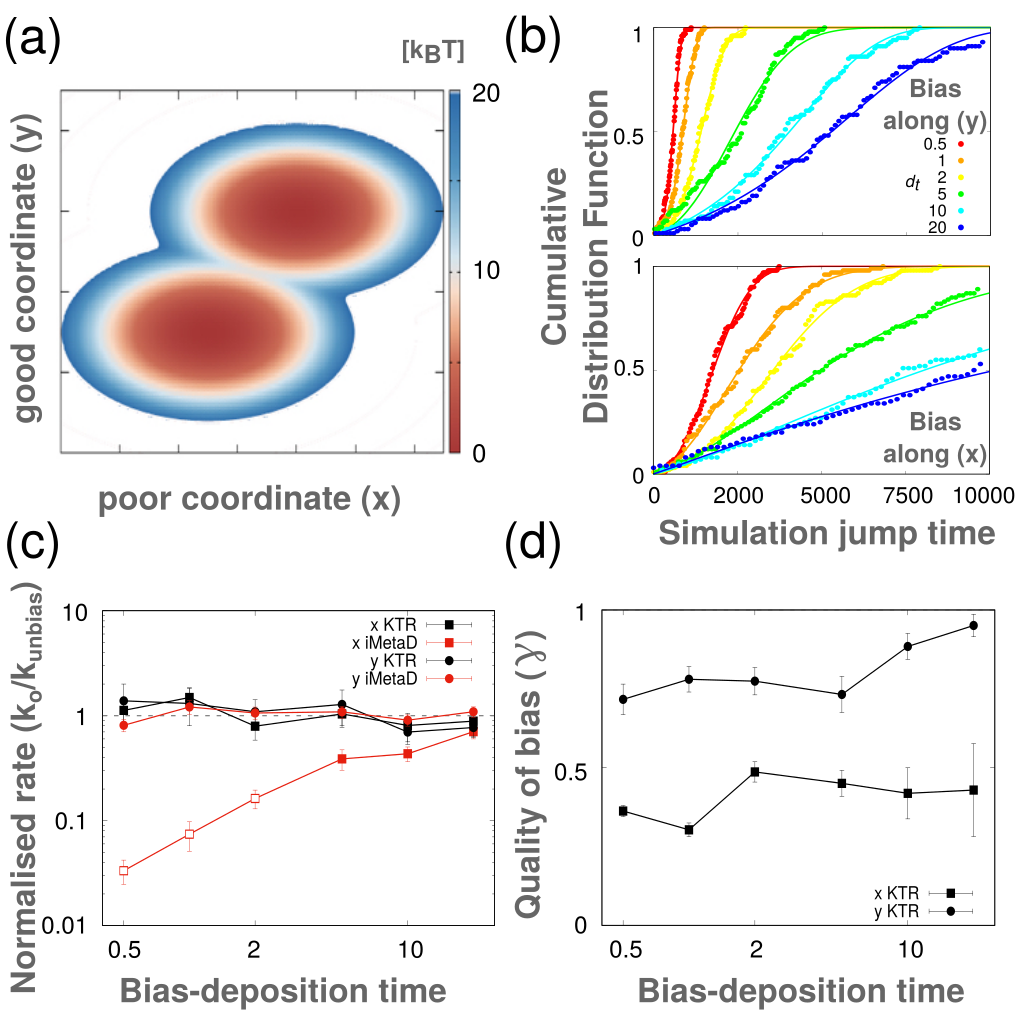}
    \caption{\textbf{MetaD Monte Carlo simulations on a 2D double-well potential.} (a) Two dimensional potential surface, where the bias is deposited along a poor ($x$) or good ($y$) CV. (b) CDF of the simulation barrier-crossing (jump) times for runs with bias along $x$ (bottom) or $y$ (top) for different bias-deposition times ($d_t$). Red to blue points go from frequent to infrequent $d_t$. Fits of the KTR theory are shown as solid lines. (c) Maximum likelihood extracted intrinsic rate ($k_0$) normalized by the true rate (calculated from unbiased simulations) as a function of the bias-deposition time for the KTR (black) and iMetaD (red) methods for biases along $x$ (squares) and $y$ (circles). Empty squares indicate cases where the KS-test failed for more than 25\% of the bootstrap trials. (d) Maximum likelihood extracted quality of bias ($\gamma$) as a function of the bias-deposition time for biases along $x$ (squares) and $y$ (circles) using the KTR method. Error bars show the standard deviation from bootstrap analysis (see the Supplementary Text).}
    \label{fig2}
\end{figure}

To study the effect of the CVs, we first tested the theory on a 2D double-well potential (Fig. \ref{fig2}a). We ran MetaD Monte Carlo simulations over a good CV ($y$) or a poor CV ($x$) (see the Supplementary Text for details) having a wide range of bias-deposition times. We started $100$ simulations from the lower well and counted a transition when the system reached the top well. Fig. \ref{fig1}b shows examples of the average $V_{\mathrm{MB}}(t)$ over the runs 
biased along $y$. In Fig. \ref{fig2}b, we show the CDFs for the simulation-jump times and their fits using $1-S(t)$ (from Eq. \ref{eq:st}) for both coordinates and the different bias-deposition times. These results show that $x$ is a poor coordinate because it does not accelerate as much the barrier-crossing events. However, in all cases there is good agreement between the theoretical and empirical CDFs. In Supplementary Fig. 1, we show CDFs for the rescaled times using iMetaD (see the Supplementary Text) together with their theoretical fits. We find that along the good CV $y$ this method works well. However, along the poor CV $x$ with frequent bias-deposition times iMetaD fails because its underlying assumptions do not hold \cite{Salvalaglio2014}. 
In Fig. \ref{fig2}c, we present the extracted rates from numerical integration of Eq. \ref{eq:st} with maximum likelihood (Eq. \ref{eq:L}) and compare them to those extracted with iMetaD by rescaling the times. We find that the KTR method estimates accurate unbiased rates, even for the challenging cases of fast deposition times and poor reaction coordinates. We note that for some of these cases (empty squares in Fig. \ref{fig2}c) the iMetaD KS-test fails, indicating that the iMetaD estimate should not be used (see Supplementary Fig. 1). Interestingly, our KTR theory enables extracting information about the quality of the biased CVs. Fig. \ref{fig2}d shows the extracted $\gamma$ as a function of the bias-deposition time for both coordinates. As expected, biasing along the poor CV leads to a lower $\gamma$. 
These results indicate that the KTR method is able to extract accurate unbiased rates and assess the quality of a CV. 

We now apply the KTR theory to study ligand unbinding from all-atom MD simulations. CDK2 is a kinase with abundant structural and pharmacological information available \cite{Kontopidis2006}. Due to its important role in the cell cycle, CDK2 has been considered a potential target for anticancer drugs \cite{Shapiro2006}. We studied the unbinding of ligand 03K (ZINC13580440), starting from PDB structure 4EK5.

\begin{figure}[ht!]
    \centering
    \includegraphics[width=8.6cm]{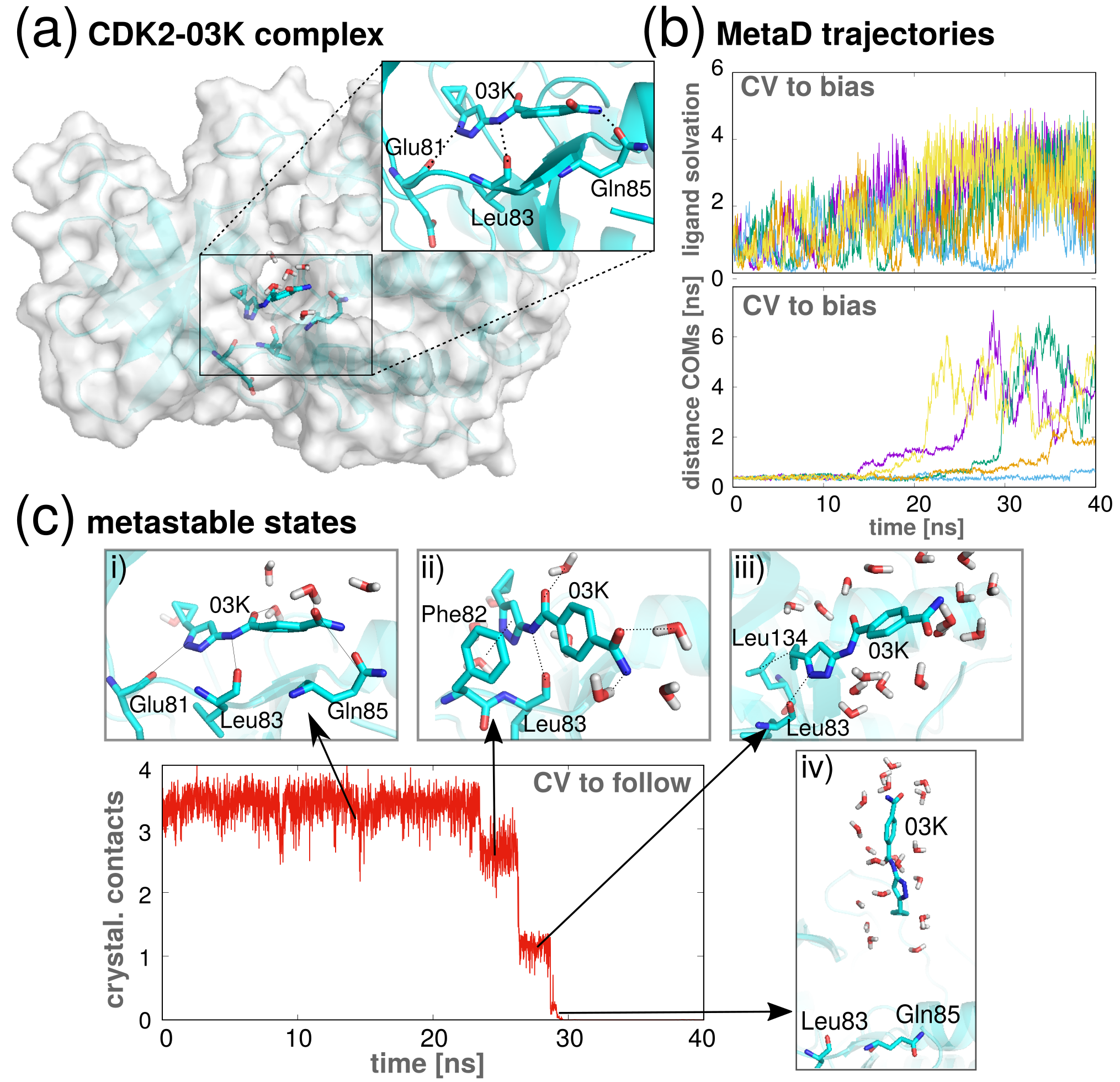}
    \caption{\textbf{CDK2-ligand unbinding biased simulations.} (a) Global view of the protein and binding pocket. We highlight the residues that interact with the ligand inside the binding pocket. (b) CDK2-ligand unbinding trajectories in five example trajectories ($d_t=10~$ps). The solvation state of the ligand (top) and the distance between the center of mass of the ligand and the center of mass of the binding pocket (bottom) are shown as a function of time. (c) Representative snapshots of the different metastable states in the path to unbinding. We follow the crystallographic contacts between the ligand and the binding pocket to define the metastable states. The arrows show the location of the snapshots along the trajectory. $i)$ Bound state: the crystallographic pose and crystallographic contacts are shown. $ii)$ and $iii)$ Before complete unbinding, two intermediate states are formed, where the ligand presents some interactions with residues within the active site. $iv)$ Unbound state: the crystallographic contacts approach to zero, and the ligand is fully solvated with no interactions inside the binding site. For our analysis, we defined the unbound state when the crystallographic-contacts CV is $<0.01$.}
    \label{fig3}
\end{figure}

We used well-tempered metadynamics \cite{Barducci2008} that modulates the height of the bias as the simulation progresses.  A global view of the protein-ligand complex and its interactions is shown in Fig. \ref{fig3}a. We estimated the rate ($k_0$) for the ligand unbinding using three sets of MetaD simulations with 50 replicas each of 10, 40 and 300 ns and bias-deposition time of 1, 10 and 100 ps, respectively (see the Supplementary Text for details). We biased two CVs which caused the ligand to unbind during the simulations in most cases. The CV1 ($w$) tracks the solvation state of the ligand and CV2 ($d$) the distance between the center of mass of the ligand and the pocket (see the Supplementary Text). In Fig. \ref{fig3}b, we show representative examples of these CVs as a function of time for five trajectories using a bias deposition time of 10 ps. To clearly identify the states, we monitor an additional CV: the number of crystallographic contacts, $c$, (see the Supplementary Text), without adding bias to it (Fig. \ref{fig3}c). When the main interactions between the ligand and the pocket are broken, the crystallographic contacts approach zero. In our analysis, we considered a transition on the final dissociation event, $i.e.,$ when CV $c$ (measuring the fraction of crystallographic contacts) is less than 0.01. However, note that the unbinding of the ligand from CDK2 involves several intermediate states before the unbound state is reached (Fig. \ref{fig3}c).

\begin{figure}[b!]
    \centering
    \includegraphics[width=8.6cm]{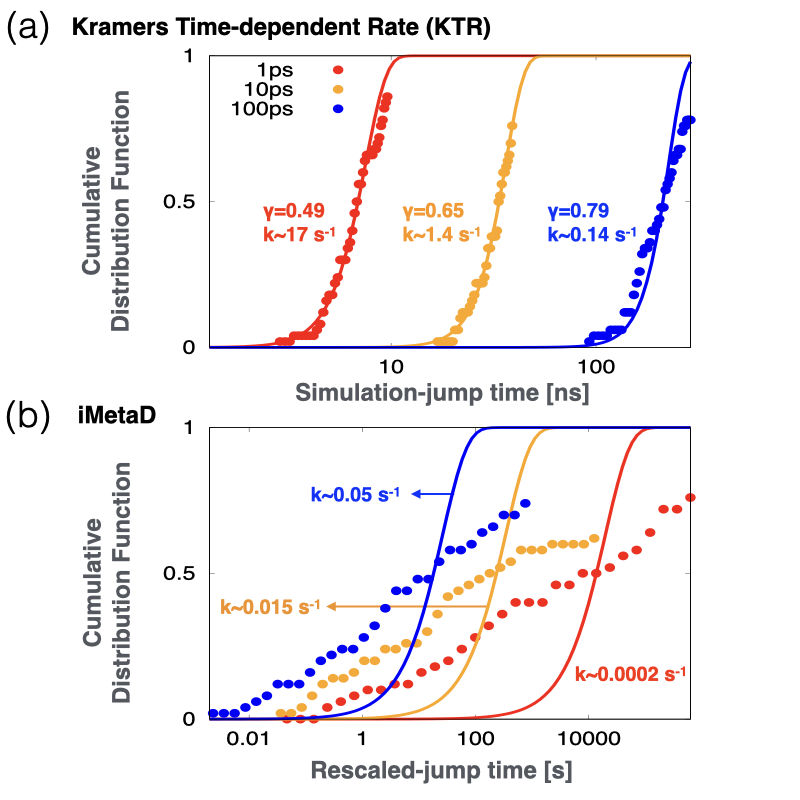}
    \caption{\textbf{Barrier-crossing statistics of CDK2-ligand unbinding.} (a) CDF of the simulation barrier-crossing (jump) times for the well-tempered metadynamics simulations for different bias-deposition times ($1$ps (red), $10$ps (orange) and $100$ps (blue)). The solid lines show the fits of the KTR method using the analytical $S(t)$ from Eq. \ref{eq:stlog}, starting from the same initial conditions to search for the optimal parameters.
    The  extracted rate $k_0$ and quality of bias $\gamma$ are shown with the same color scheme. The KS-tests pass for all bias-deposition time setups.
    (b) CDF of the iMetaD rescaled times from the same simulations. Attempted fits are shown as solid lines, together with extracted rates. For these cases the KS-test fails.}
    \label{fig4}
\end{figure}

We found that, for all simulation setups, the time dependence of the average maximum bias $V_{\mathrm{MB}}(t)$ was well fitted to a logarithmic function  $f(t)=a\log(1+b\,t)$ (see Supplementary Fig. 2). This allowed us to use the analytic expression of $S(t)$ in Eq. \ref{eq:stlog} (with fitted parameters $a$ and $b$ fixed) to extract $\gamma$ and $k_0$, and to perform KS-tests against the analytic S(t) (Eq. \ref{eq:stlog}).
In Fig. \ref{fig4}a, we show the empirical CDFs for the simulation barrier-crossing times for the different setups together with the fits of the KTR formalism. 
The estimated $\gamma$ and $k_0$, and their errors (extracted from bootstrap analysis of the passed KS-trials) are shown in Supplementary Fig. 3.  As expected, $\gamma$ increases for slower deposition times; however, even for the best case its median is around 0.78, indicating that the bias setup and CVs are not perfect. In comparison to the experimental rate $k_{exp}=0.26 \pm 0.05 s^{-1}$\cite{Dunbar2013}, the extracted rates approach the experimental one as the quality of bias increases. For the longest bias-deposition time, the estimate is $k_0=0.14 \pm 0.03 s^{-1}$, which is on the same order of magnitude as the experiment. We consider this a successful result since we are using simulations of maximum length 300 ns to estimate a value that is on the order of seconds. Moreover, the KTR method proves extremely valuable when predicting the underlying statistics of barrier-crossing times in comparison to attempted CDF-fits using the iMetaD rescaled times (Fig. \ref{fig4}b)  \cite{Tiwary2013,Salvalaglio2014}.


Inspired by the methods from the force spectroscopy community \cite{Hummer2003,Dudko2006,Cossio2016} that calculate barrier-crossing rates induced by forces acting on single molecules, in this work, we used Kramers' theory to calculate the time-dependent transition rates and survival probabilities from time-dependent biased simulations. Here, we have used examples from MetaD simulations, however, the KTR method is general for any time-dependent biased simulation along some CVs, such as, adaptive biasing force \cite{Darve2008,Henin2010}, adiabatic bias MD \cite{Marchi1999,Paci1999}, adaptively biased MD \cite{Babin2008,Babin2009}, among others. 
Importantly, our method not only enables estimating the unbiased intrinsic rate but also provides a measure for the effectiveness of the added bias to accelerate the transition (related to the quality of the CVs). This overcomes severe limitations encountered with previous approaches where the bias had to be deposited very infrequently over ideal CVs. 

There are several points of the KTR theory that can be improved in future work. For example, there could be cases where multiple values of $k_0$ and $\gamma$ fit equally well the CDFs, and additional restrictions over these parameters might be required. Automatized methods to determine when the barrier-crossing occurs might be helpful. Moreover, Kramers' high-barrier approximation or the quasi-adiabatic assumption might breakdown for extremely large biases or deposition times. A generalization to multiple-basins and dimensions would also be useful.

P.C. was supported by MinCiencias, University of Antioquia (Colombia), the Max Planck Society (Germany), and Flatiron Institute, a division of the Simons Foundation (USA). The authors thank Attila Szabo, Erik Thiede and Marylou Gabri\'e for useful discussions, as well as David Silva for code optimization advice. Calculations were performed on the GENCI-IDRIS French national supercomputing facility, under grant number A0090811069.
K.P-R. and H.V. contributed equally to this work.


\bibliography{biblio}

%

%% file: text_SI.tex
\setcounter{figure}{0}  
\setcounter{equation}{0}

\title{Supplementary Information: Transition rates, survival probabilities, and quality of bias from time-dependent biased simulations}

\author{Karen Palacio-Rodriguez$^{1,2}$}
\author{Hadrien Vroylandt$^{3}$}
\author{Lukas S. Stelzl$^{4,5,6,7}$}
\author{Fabio Pietrucci$^{1}$}
\author{Gerhard Hummer$^{7,8}$}
\author{Pilar Cossio$^{2,7,9}$}
\email{pcossio@flatironinstitute.org}
\affiliation{$^{1}$Sorbonne Université, Institut de Minéralogie, de Physique des Matériaux et de Cosmochimie, IMPMC, F-75005 Paris, France}
\affiliation{$^{2}$Biophysics of Tropical Diseases Max Planck Tandem Group, University of Antioquia, Medellín, Colombia}
\affiliation{$^{3}$Sorbonne Université, Institut des sciences du calcul et des données, ISCD, F-75005 Paris, France}
\affiliation{$^{4}$Faculty of Biology, Johannes Gutenberg University Mainz, Gresemundweg 2, 55128 Mainz, Germany}
\affiliation{$^{5}$KOMET 1, Institute of Physics, Johannes Gutenberg University  Mainz, 55099 Mainz, Germany}
\affiliation{$^{6}$Institute of Molecular Biology (IMB), 55128 Mainz, Germany}
\affiliation{$^{7}$Department of Theoretical Biophysics, Max Planck Institute of Biophysics, Max-von-Laue Straße 3, 60438 Frankfurt am Main, Germany}
\affiliation{$^{8}$Institute for Biophysics, Goethe University Frankfurt, 60438 Frankfurt am Main, Germany}
\affiliation{$^{9}$Center for Computational Mathematics, Flatiron Institute, New York, USA}



\maketitle

\section{Supplementary Text}

\subsection{Rate-acceleration factor: quality of bias $\gamma$}

Consider the diffusive escape from a 2D double-well
potential over a cusp-like barrier. The potential is given by
\begin{equation}
    V(x,y)=
    \begin{cases}
			\frac{y^2}{2}+\frac{(x-ay)^2}{2}, & \text{for $y \leq 1$}~,\\
            -\infty, & \text{for $y > 1$}~.
	\end{cases}
\end{equation}

The minimum of the potential well is at $x = y = 0$, and the barrier for escape at $x = a$ and $y = 1$. The barrier height is $\Delta G^\ddagger_0 = \frac{1}{2}$. We consider diffusion on this potential with a uniform and isotropic diffusion coefficient at an inverse temperature $\beta \gg 1$, $i.e.$, in the high-barrier limit. Then, for $a \approx 0$, $y$ is a good CV because it reports on the approach to the barrier, whereas $x$ is a poor CV because it reports on motions orthogonal to the escape over the barrier.

Now, consider that metadynamics is performed with $x$ as the chosen CV. We assume that bias deposition is very slow such that we have quasi-equilibrium conditions. Then, metadynamics flattens the potential of mean force (PMF) along $x$ up to a preset level $\Delta G$. The PMF along $x$ is defined as 
\begin{equation}
    e^{-\beta g(x)}=\int^1_{-\infty} dy e^{-\beta V(x,y)}~,
\end{equation}
up to an additive constant, chosen such that the minimum is at a PMF value of zero. 
For small $a$ and $x$, the PMF along $x$ can be approximated as
\begin{equation}
  g(x) \approx  \frac{x^2}{2(1+a^2)}~.
\end{equation}

For a given metadynamics bias level $\Delta G$, the bias acts on the range $|x| < \sqrt{2(1+a^2)\Delta G}$. The combined potential including the metadynamics bias is then
\begin{equation}
\begin{split}
        U(x,y|\Delta G) = 
        \begin{cases}
			V(x,y)-g(x), & \text{for $|x| < \sqrt{2(1+a^2)\Delta G}$}\\
            V(x,y)-\Delta G, & \text{otherwise}~.
        \end{cases}
\end{split}
\end{equation}

The potential $U(x,y)$ has a minimum in the bound well for $|a| < 1/\sqrt{2\Delta G -1}$. The minimum is located at $x = \sqrt{2(1+a^2)\Delta G}$ and $y = a\sqrt{2\Delta G/(1+a^2)}$. The lowest energy barrier to escape on the combined potential is
located at $x = \sqrt{2(1+a^2)\Delta G}$ and $y = 1$. For $|a| < 1/\sqrt{2\Delta G -1}$, the height of the barrier on the potential $U(x,y)$ including the metadynamics bias is
\begin{equation}
    \Delta G^\ddagger = \frac{1+a[a-\sqrt{8(1+a^2)\Delta G} + 2a\Delta G]}{2}~.
\end{equation}

Now if metadynamics has reached a level of $\Delta G = \Delta G^\ddagger_0 = 1/2$, $i.e.$, the level of the barrier in the potential $V(x,y)$, the potential well is nominally filled. However, as the above calculation shows, for $|a| < 1/\sqrt{2\Delta G - 1}$ a barrier $\Delta G^\ddagger_0$ remains. For small $a$ and $\Delta G = \Delta G^\ddagger_0 = 1/2$, the height of this remaining barrier is
\begin{equation}
    \Delta G^\ddagger \approx \frac{1}{2}-a~.
\end{equation}

If one now uses the Kramers' approximation for the rate of escape from the potential well without bias (Eq.~1 Main Text),
\begin{equation*}
    k_0 = k_{\mathrm{pre}}e^{-\beta\Delta G^\ddagger_0}~.
\end{equation*}
and with a bias potential filled up along $x$ to a height of $\Delta G = \Delta G^\ddagger_0$, then the rate accelerates to
\begin{equation}
    k_\mathrm{meta}= k_{\mathrm{pre}}e^{-\beta  \Delta G^\ddagger } \approx k_{\mathrm{pre}}e^{-\beta\Delta G^\ddagger_0(1-2a)}~.
\end{equation}
This means that we have not gotten the full boost from metadynamics. In the Main Text, we correct for this reduced boost by the factor $\gamma$. For the problem here, this factor is obtained from
\begin{equation}
    k_\mathrm{meta} \approx k_{\mathrm{pre}}e^{-\beta\Delta G^\ddagger_0(1-2a)} = k_{\mathrm{pre}}e^{-\beta\Delta G^\ddagger_0(1-\gamma)}~.
\end{equation}

Therefore, for the escape from this double well and $a \approx 0$, the quality of bias is thus
\begin{equation}
    \gamma \approx 2a~.
\end{equation}

In other words, for $a \rightarrow 0$, when $x$ is orthogonal to the escape flux, we have no acceleration from metadynamics, and from $0 < a \ll 1$, we have only a much smaller acceleration than what would be achieved by biasing along a well-chosen CV.

If instead $y$ is chosen as the CV, metadynamics flattens the PMF along $y$. The PMF along $y$ is given, up to a constant, by
\begin{equation}
    G(y)=\frac{y^2}{2}~.
\end{equation}

The combined potential is
\begin{equation}
    U(x,y|\Delta G)=
    \begin{cases}
			V(x,y)-G(y), & \text{for $|y| < \sqrt{2\Delta G}$}\\
            V(x,y)-\Delta G, & \text{otherwise}~.
	\end{cases}
\end{equation}
For $\Delta G = \Delta G^\ddagger_0 = 1/2$, the barrier to the exit at $x = a$ and $y = 1$ vanishes. Therefore, the choice of $y$ as a CV results in a boost factor $\gamma = 1$ for all values of $a$.

\subsection{Analytical $S(t)$ for a linear time-dependent bias}

Assuming a linear time dependence of the bias $V_{\mathrm{MB}}(t) = a\,t$, and using Eq. 4. (Main Text), we derived the following analytical expression for the survival probability
\begin{equation}
    S(t) =  \exp\left(\frac{k}{\beta\gamma\,a}\left(1-e^{\beta\gamma\,a\,t}\right)\right)~,
    \label{eq:stlinear}
\end{equation}
where $\gamma$ and $k$ are the quality of bias and intrinsic transition rate, respectively.

\subsection{Maximization of the likelihood function}

Combining the expressions of the likelihood from Eq.~6 (Main Text) and the survival function from Eq.~4 (Main Text), we obtain an expression for the log-likelihood

\begin{align}
    \ln \mathcal{L}(k_0,\gamma) =-k_0\sum_{i }^{M+N}  \int_0^{t_i} e^{\beta\gamma V_{\mathrm{MB}}(t')}dt' \nonumber\\ +M\ln k_0 +\sum_{i \in \mathrm{events}} ^M  \beta \gamma V_{\mathrm{MB}}(t_i)~,
    \label{eq:logL}
\end{align}

where $M$ is the number of events and $N$ the number of non-events. Our aim is to extract parameters $k^*_0$ and $\gamma^*$ that maximize this function. The derivative with respect to $k_0$  of the log-likelihood is 
\begin{equation}
 \frac{\partial }{\partial k_0}  \ln \mathcal{L}(k_0,\gamma) = -\sum_{i }^{M+N}  \int_0^{t_i} e^{\beta\gamma V_{\mathrm{MB}}(t')}dt'+\frac{M}{k_0}~.
\end{equation}
 To solve for the optimal intrinsic rate, we set the previous derivative to zero, and find it as a function of $\gamma$
\begin{equation}
    k^{*}_0(\gamma) = \frac{M}{\sum_{i}^{M+N} \int_0^{t_i} e^{\beta\gamma V_{\mathrm{MB}}(t')}dt'} ~.
\end{equation}
Using this function, we search for the optimal parameters $\gamma^*$ and $k^*_0 = k^{*}_0(\gamma^*)$ through numerical maximization of $ \ln\mathcal{L}(k^{*}_0(\gamma),\gamma)$.

\subsection{Rate from iMetaD}

In MetaD, the kinetic rates of the system are altered in a complex way because of the acceleration due to the time-dependent bias, constructed as a sum of repulsive Gaussian-shaped functions deposited at regular time intervals within a low-dimensional CV space. Different approaches have been designed to recover unbiased kinetic rates from MetaD using Markov state models \cite{Marinelli2009,Donati2018}. iMetaD, proposed by \citeauthor{Tiwary2013} \cite{Tiwary2013}, is a simple and commonly applied approach that starts by repeating a series of identical MetaD simulations where Gaussians are deposited relatively slowly, in order to reduce the probability of biasing the transition state region. Under these conditions, ideas from \citeauthor{Grubmuller1995a} \cite{Grubmuller1995a} and \citeauthor{Voter1997a} \cite{Voter1997a} based on TST are exploited to recover the unbiased dynamics of the system by rescaling the simulation time with an acceleration factor correcting for the bias. In practice, we rescale the times from MetaD by \cite{Casasnovas2017}
\begin{equation}
    t_\mathrm{res}=\sum_{i=1}^{N_t} \mathrm{d}t \:e^{\beta V^B_i},
    \label{eq:t-inf-pract}
\end{equation}
where $\mathrm{d}t$ is the time step, $i$ is the step number, $N_t$ is the total number of steps until an event occurred or the simulation stopped, $V^B_i$ is the instantaneous MetaD biasing potential at step $i$ and $\beta=1/(k_BT)$ with $k_B$ the Boltzmann's constant and $T$ the absolute temperature.

\subsection{Cumulative distribution function and Kolmogorov-Smirnov test} 

The Kolmogorov-Smirnov test (KS-test) is a non-parametric test that enables comparing the similarity between one-dimensional probability distributions. 

In iMetaD, the reliability of the rescaled escape (residence) times is quantitatively assessed by comparing their empirical cumulative distribution function (ECDF) to the theoretical cumulative distribution function (TCDF) of a homogeneous Poisson process using a two-sample KS test \cite{Salvalaglio2014}. A $p$-value threshold of 0.05 is typically used \cite{Salvalaglio2014}. The ECDF is fitted to the TCDF of a homogeneous Poisson process, $P(t) = 1-\exp({-t/\tau})$. 
Following \citeauthor{Salvalaglio2014} \cite{Salvalaglio2014}, a TCDF is built from a large number of sample times (\textit{e.g.}, $10^6$) randomly generated according to $P(t)$. Then, a two sample KS test is applied to assess the similarity between ECDF and TCDF, with the null hypothesis that the two sampled distributions share the same underlying distribution. If the $p$-value from the KS-test is higher that 0.05, ignoring the fact that $\tau$ has been fitted, then the null hypothesis is accepted and a Poissonian process is considered appropriate to describe the statistics of the process. The rate coefficient is then calculated as the inverse of the average time from the fit of the ECDF, $k_{\mathrm{iMetaD}}=1/\tau$. 

In contrast, the KTR method does not rescaled the times and therefore, is not constrained to assume Poissonian statistics. When the analytical form of $V_{\mathrm{MB}}(t)$ is known ($e.g.,$ linear or logarithmic time-dependency), the survival probability ($S(t)$) can be extracted from Eq. 4 (Main Text), and the TCDF is given by $1 - S(t)$. One can perform the KS-test as described above using the new TCDFs and comparing it to the ECDF of the simulation-jump times, extracting a p-value that indicates if the KS-test passed or not. 

We adapted a Matlab code provided by \citeauthor{Salvalaglio2014} \cite{Salvalaglio2014} to python3.6 for the calculation of $\tau$ using iMetaD and the KS-test (for both methods, iMetaD and KTR). The histograms of the CDFs are built using the number of bins equal the total number of simulations (this was tested empirically to obtain the best fits of the CDF for the 2D system). We use logarithmic scale to create the histograms for iMetaD and a linear scale for KTR.

\subsection{Practical considerations}

\textbf{Bootstrap analysis:}

To estimate the errors of the extracted parameters, we performed a bootstrapping analysis. The number of bootstrapping samples is 100 for all systems. We use the distributions of each parameter from the bootstrap samples to estimate its error (Fig. 2, Fig. 4a and, SI Fig. 3).

\textbf{Code availability:}

The codes to fit analytically or numerically $V_{MB}(t)$ and the survival probabilities are available on GitHub: \url{https://github.com/kpalaciorodr/KTR}

\subsection{2D double-well potential}
To analyse the dependence of the results on the quality of the CVs, we used a 2D double-well potential $U(x) = -k_BT \ln[e^{(-20\cdot(x-0.2)^2 -100\cdot(y-0.2)^2)} + e^{(-20\cdot(x-0.8)^2-100\cdot(y-0.8)^2)}]$ (similar to that in ref. \citenum{Rosta2015}). For this potential, the projection of the free energy surface along the $x$ coordinate leads to an underestimation of the barrier between the wells, while the projection along the $y$ coordinate represents faithfully the underlying 2D barrier of around $8~k_BT$ (see Fig. 2a). The Monte Carlo (MC) step was performed randomly from a uniform distribution for each CV in the range [-0.005, 0.005] around the current point inside a grid between -0.4 and 1.4 for $x$ and $y$. These trial moves were accepted according to the Metropolis criterion. A bias height of 0.04 and a bias width of 0.04 were used. To asses the effect of the CV, the simulations were performed adding bias only along a single coordinate (\textit{i.e.}, either the $x$ or $y$ coordinate, in independent simulations). The bias deposition time was varied and 100 simulations were launched for each CV. To obtain the reference rate coefficient $k_{unbias}$ for the double-well potential, we performed 1000 unbiased runs with $3\times10^{7}$ MC steps. We calculated $k_{unbias}= M/(\sum_i t_{i} + \sum_j  t_{j})$ where $M$ is the total number of events, $t_i$ is the escape time observed in run $i$, and $t_j$ is the total simulation time for run $j$ that did not have a transition (for the unbiased case around 30\% of non-events).

\subsection{CDK2-ligand unbinding simulations:}

Well-tempered MetaD simulations \cite{Barducci2008} were carried out using the GROMACS 2019.4 program \cite{Berendsen1995, Abraham2015a} patched with PLUMED 2.5.3 \cite{Tribello2014}. The complex was solvated with a cubic water box, centered at the geometric center of the complex with at least 2.0 nm between any two periodic images. The AMBER99SB-ILDN \cite{Lindorff-Larsen2010} force field was used to model the system with the TIP3P water model \cite{Jorgensen1983}.  The ligand was parameterized using antechamber \cite{Wang2006} with GAFF \cite{Wang2004a}. The parameters found were converted into GROMACS format using ACPYPE \cite{SousadaSilva2012}. A minimization was done with the steepest descent algorithm and stopped when the maximum force was $\leq$ 1000 kJ/mol$\cdot$nm. Periodic boundary conditions were considered. We used the leapfrog algorithm to propagate the equations of motion and the nonbonding interactions were calculated using a PME scheme with a 1.0 nm cutoff for the part in real space. We performed a 100 ps equilibration in an NVT ensemble using the the velocity rescaling thermostat \cite{bussi2007} followed by a 100 ps equilibration in an NPT ensemble using Parrinello-Rahman barostat \cite{parrinello1981} with a time step of 2 fs. The MD production was performed without restrains, with a time step of 2 fs in an NPT ensemble at 300.15 K and 1 atm. 

We chose two CVs for the well-tempered MetaD simulations. The first CV was the solvation state of the ligand ($w$), calculated as the coordination number between two groups
\begin{equation}
    w = \sum_{i \in A} \sum_{i \in B} w_{ij},
    \label{eq:w_CV}
\end{equation}
with
\begin{equation}
    w_{ij} = \frac{1-\Big(\frac{r_{ij}-d_0}{r_0}\Big)^n}{1-\Big(\frac{r_{ij}-d_0}{r_0}\Big)^m},
    \label{eq:wij}
\end{equation}
where $d_0=0$, $r_0=0.3$, $n=6$ and $m=10$. In the sum of Eq. \ref{eq:w_CV}, group A is the center of mass (COM) of the ligand and group B are the oxygen atoms of all water molecules at a distance shorter than 5 \AA \space from the pocket. The second CV was the distance between  binding pocket and ligand ($d$). We define $d$ as the distance between the COM of the heavy atoms in the ligand and the COM of the $\alpha$-carbons in the binding pocket, i.e., at a distance of 5 \AA \space from the ligand in the binding pose.  

Well-tempered MetaD \cite{Barducci2008} is performed with an initial Gaussian height of 1.5 kJ/mol. The width ($\sigma$) of the Gaussians was $\sigma_w=0.13$ and $\sigma_d=0.02$ nm, for the $w$ and $d$ CVs, respectively. We used a bias factor of 15. We performed independent simulations where Gaussians were deposited every 1, 10 and 100 ps with total simulation time of 10, 40 and 300 ns, respectively. 50 simulations per Gaussian deposition time were performed.

To monitor the escape of the ligand from the binding pocket, we followed (without biasing it) the evolution of an additional CV, the crystallographic contacts ($c$). Using $c$ we can distinguish easily between the bound and unbound state of the ligand. To define $c$ we use an equation analogous to Eq. \ref{eq:wij}, involving the atoms responsible for the main interactions between the ligand and the binding pocket, $i.e.,$ group A are the nitrogen atoms of the ligand that form hydrogen bonds -- in the binding pose -- with the atoms in the group B: O-Glu81, O-Leu83 and O-Gln85. These interactions are shown in Fig. 3a (Main Text).

\clearpage

\section{Supplementary Figures}

\begin{figure}[htb!]
    \centering
    \includegraphics[width=0.45\textwidth]{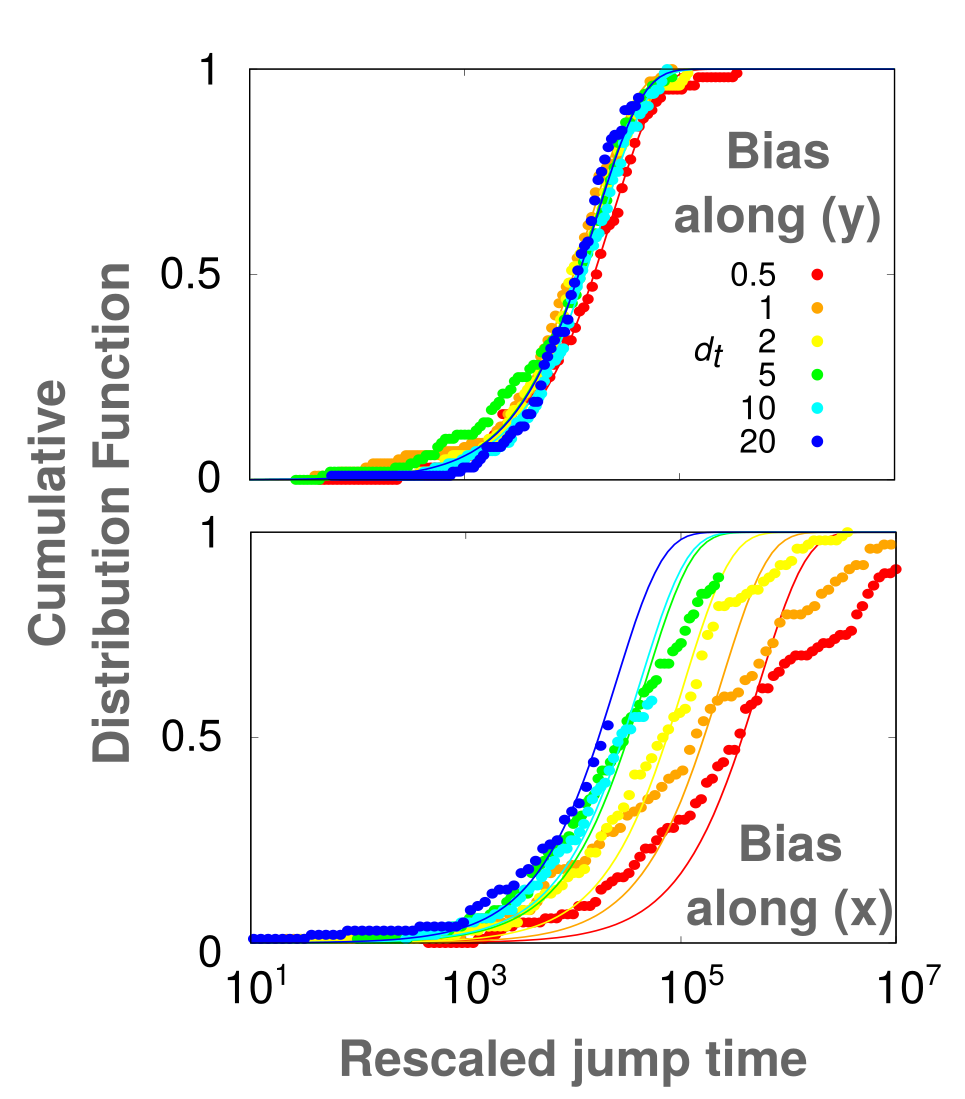}
    \caption{\textbf{CDFs of rescaled times and fits to Poisson distributions along x and y.}
    CDFs of the rescaled barrier-crossing (jump) times from iMetaD. Runs with bias along $x$ (bottom) or $y$ (top) for different bias-deposition times ($d_t$). Red to blue points go from frequent to infrequent $d_t$. Fits to the respective Poisson distribution are shown as solid lines \cite{Salvalaglio2014}.}
    \label{SI_fig1}
\end{figure}

\begin{figure}[htb!]
    \centering
    \includegraphics[width=0.4\textwidth]{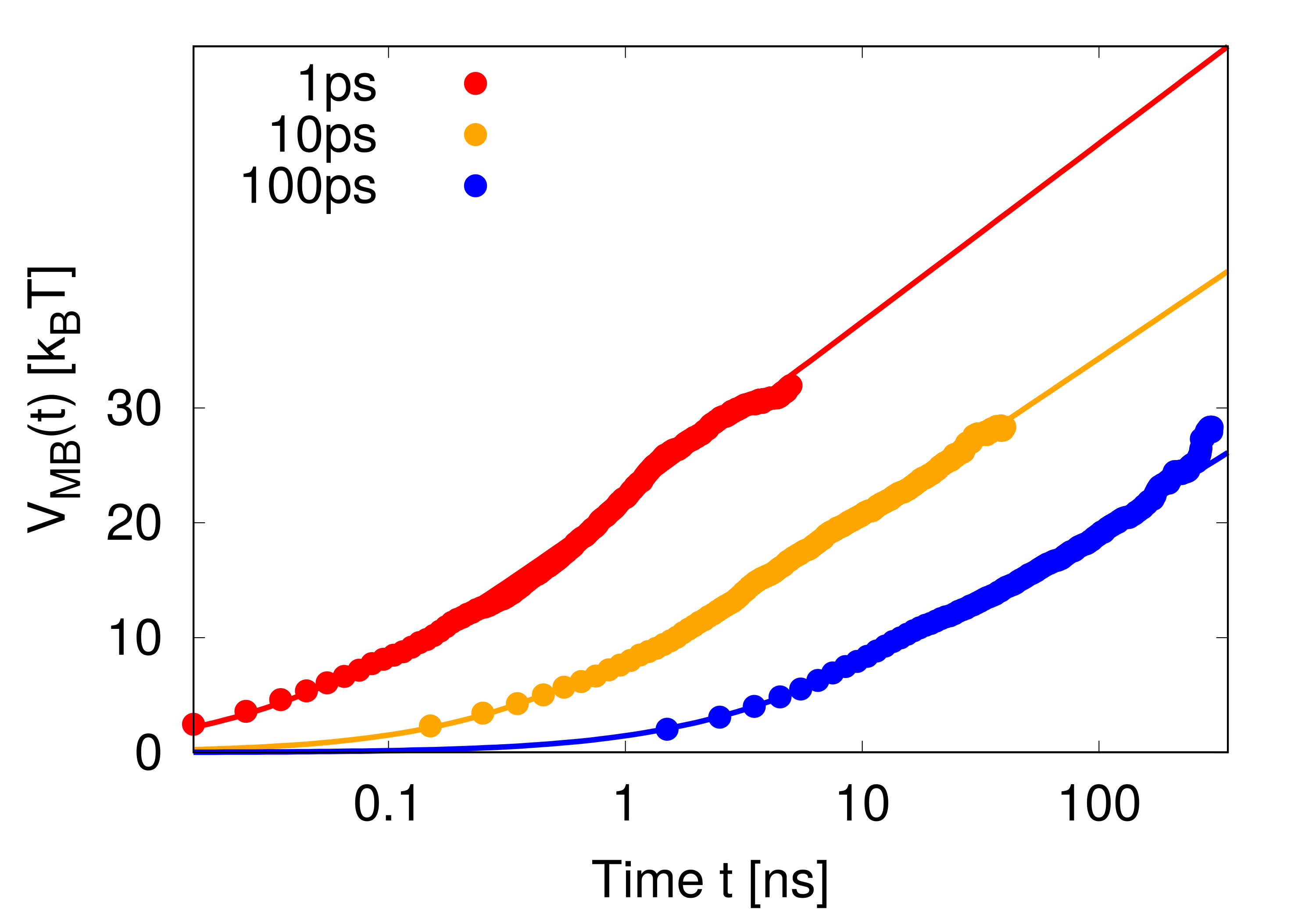}
    \caption{\textbf{Fits of $V_{MB}(t)$ for the CDK2-ligand unbinding simulations using a logarithmic function.} Time-dependent maximum bias $V_{\mathrm{MB}}(t)$ (see Main Text Eq. 2) for the CDK2-ligand unbinding simulations. Solid lines show the fits to a logarithmic function $f(t,a,b)=a\log(1+b\,t)$. The fit for the longest bias-deposition time was up to $200$ns. }
    \label{SI_fig2}
\end{figure}

\begin{figure}[htb!]
    \centering
    \includegraphics[width=0.4\textwidth]{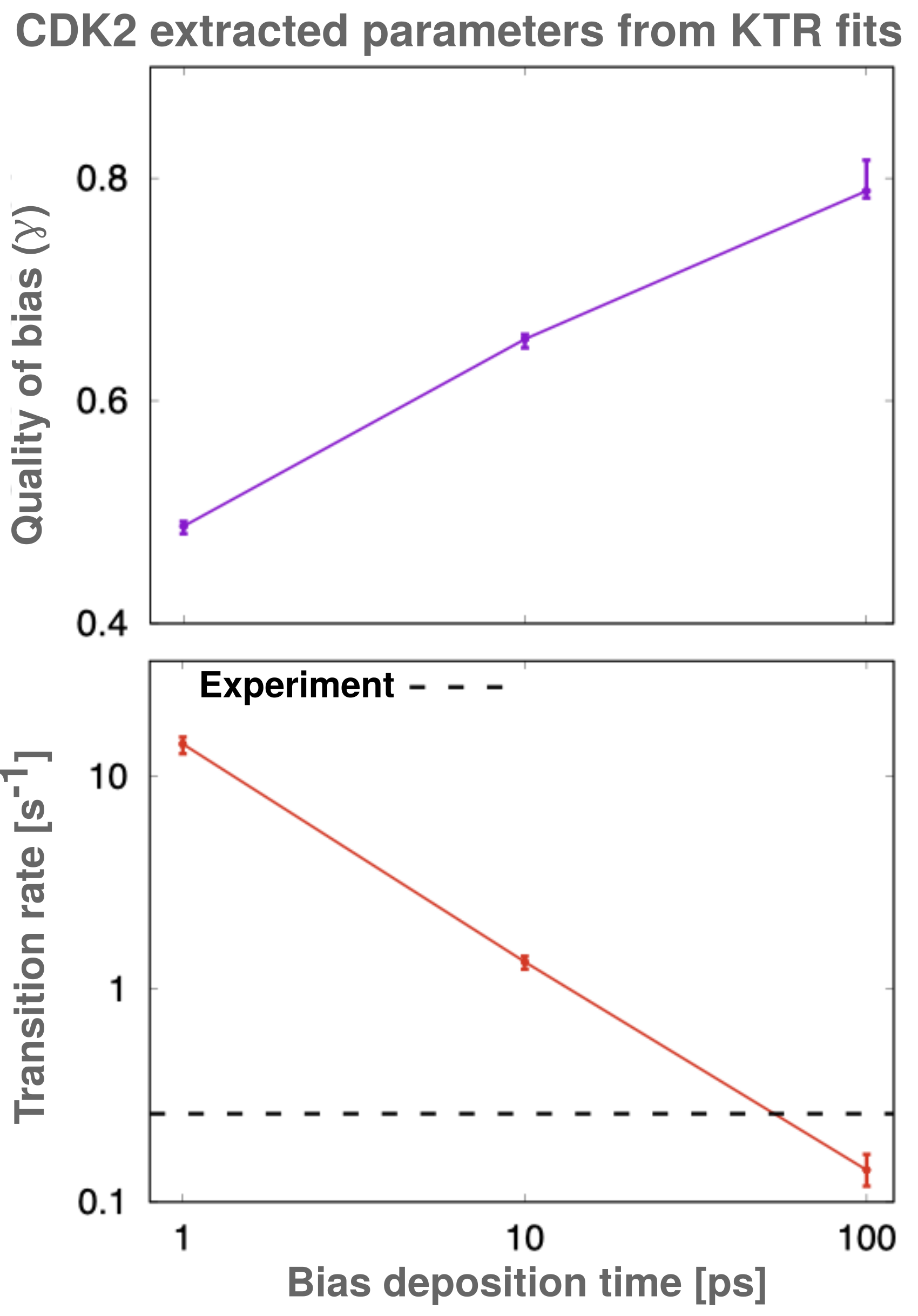}
    \caption{\textbf{Extracted parameters from CDK2-ligand unbinding simulations using the KTR theory.} (top) Quality of bias $\gamma$ and (bottom) transition rate $k_0$ extracted using the KTR method for the CDK2-ligand unbinding simulations with different bias-deposition times. The results show the median, and credible interval calculated at 25\% and 75\% for the distribution of trials that passed the KS-test from bootstrap analysis (which were more than 50\% of the trails for all cases). The experimental value of the transition rate is shown as a dashed black line.}
    \label{SI_fig3}
\end{figure}



%

%% file: ms.bbl
\begin{thebibliography}{75}%
\makeatletter
\providecommand \@ifxundefined [1]{%
 \@ifx{#1\undefined}
}%
\providecommand \@ifnum [1]{%
 \ifnum #1\expandafter \@firstoftwo
 \else \expandafter \@secondoftwo
 \fi
}%
\providecommand \@ifx [1]{%
 \ifx #1\expandafter \@firstoftwo
 \else \expandafter \@secondoftwo
 \fi
}%
\providecommand \natexlab [1]{#1}%
\providecommand \enquote  [1]{``#1''}%
\providecommand \bibnamefont  [1]{#1}%
\providecommand \bibfnamefont [1]{#1}%
\providecommand \citenamefont [1]{#1}%
\providecommand \href@noop [0]{\@secondoftwo}%
\providecommand \href [0]{\begingroup \@sanitize@url \@href}%
\providecommand \@href[1]{\@@startlink{#1}\@@href}%
\providecommand \@@href[1]{\endgroup#1\@@endlink}%
\providecommand \@sanitize@url [0]{\catcode `\\12\catcode `\$12\catcode
  `\&12\catcode `\#12\catcode `\^12\catcode `\_12\catcode `\%12\relax}%
\providecommand \@@startlink[1]{}%
\providecommand \@@endlink[0]{}%
\providecommand \url  [0]{\begingroup\@sanitize@url \@url }%
\providecommand \@url [1]{\endgroup\@href {#1}{\urlprefix }}%
\providecommand \urlprefix  [0]{URL }%
\providecommand \Eprint [0]{\href }%
\providecommand \doibase [0]{https://doi.org/}%
\providecommand \selectlanguage [0]{\@gobble}%
\providecommand \bibinfo  [0]{\@secondoftwo}%
\providecommand \bibfield  [0]{\@secondoftwo}%
\providecommand \translation [1]{[#1]}%
\providecommand \BibitemOpen [0]{}%
\providecommand \bibitemStop [0]{}%
\providecommand \bibitemNoStop [0]{.\EOS\space}%
\providecommand \EOS [0]{\spacefactor3000\relax}%
\providecommand \BibitemShut  [1]{\csname bibitem#1\endcsname}%
\let\auto@bib@innerbib\@empty
\bibitem [{\citenamefont {Holenz}\ and\ \citenamefont
  {Stoy}(2019)}]{Holenz2019}%
  \BibitemOpen
  \bibfield  {author} {\bibinfo {author} {\bibfnamefont {J.}~\bibnamefont
  {Holenz}}\ and\ \bibinfo {author} {\bibfnamefont {P.}~\bibnamefont {Stoy}},\
  }\bibfield  {title} {\bibinfo {title} {Advances in lead generation},\
  }\href@noop {} {\bibfield  {journal} {\bibinfo  {journal} {Bioorganic \&
  medicinal chemistry letters}\ }\textbf {\bibinfo {volume} {29}},\ \bibinfo
  {pages} {517} (\bibinfo {year} {2019})}\BibitemShut {NoStop}%
\bibitem [{\citenamefont {Rognan}(2017)}]{Rognan2017a}%
  \BibitemOpen
  \bibfield  {author} {\bibinfo {author} {\bibfnamefont {D.}~\bibnamefont
  {Rognan}},\ }\bibfield  {title} {\bibinfo {title} {The impact of in silico
  screening in the discovery of novel and safer drug candidates},\ }\href@noop
  {} {\bibfield  {journal} {\bibinfo  {journal} {Pharmacology \& therapeutics}\
  }\textbf {\bibinfo {volume} {175}},\ \bibinfo {pages} {47} (\bibinfo {year}
  {2017})}\BibitemShut {NoStop}%
\bibitem [{\citenamefont {Talele}\ \emph {et~al.}(2010)\citenamefont {Talele},
  \citenamefont {Khedkar},\ and\ \citenamefont {Rigby}}]{Talele2010}%
  \BibitemOpen
  \bibfield  {author} {\bibinfo {author} {\bibfnamefont {T.~T.}\ \bibnamefont
  {Talele}}, \bibinfo {author} {\bibfnamefont {S.~A.}\ \bibnamefont
  {Khedkar}},\ and\ \bibinfo {author} {\bibfnamefont {A.~C.}\ \bibnamefont
  {Rigby}},\ }\bibfield  {title} {\bibinfo {title} {Successful applications of
  computer aided drug discovery: moving drugs from concept to the clinic},\
  }\href@noop {} {\bibfield  {journal} {\bibinfo  {journal} {Current topics in
  medicinal chemistry}\ }\textbf {\bibinfo {volume} {10}},\ \bibinfo {pages}
  {127} (\bibinfo {year} {2010})}\BibitemShut {NoStop}%
\bibitem [{\citenamefont {Jorgensen}(2004)}]{Jorgensen2004}%
  \BibitemOpen
  \bibfield  {author} {\bibinfo {author} {\bibfnamefont {W.~L.}\ \bibnamefont
  {Jorgensen}},\ }\bibfield  {title} {\bibinfo {title} {The many roles of
  computation in drug discovery},\ }\href@noop {} {\bibfield  {journal}
  {\bibinfo  {journal} {Science}\ }\textbf {\bibinfo {volume} {303}},\ \bibinfo
  {pages} {1813} (\bibinfo {year} {2004})}\BibitemShut {NoStop}%
\bibitem [{\citenamefont {Bernetti}\ \emph {et~al.}(2019)\citenamefont
  {Bernetti}, \citenamefont {Masetti}, \citenamefont {Rocchia},\ and\
  \citenamefont {Cavalli}}]{Bernetti2019}%
  \BibitemOpen
  \bibfield  {author} {\bibinfo {author} {\bibfnamefont {M.}~\bibnamefont
  {Bernetti}}, \bibinfo {author} {\bibfnamefont {M.}~\bibnamefont {Masetti}},
  \bibinfo {author} {\bibfnamefont {W.}~\bibnamefont {Rocchia}},\ and\ \bibinfo
  {author} {\bibfnamefont {A.}~\bibnamefont {Cavalli}},\ }\bibfield  {title}
  {\bibinfo {title} {Kinetics of drug binding and residence time},\ }\href@noop
  {} {\bibfield  {journal} {\bibinfo  {journal} {Annual review of physical
  chemistry}\ }\textbf {\bibinfo {volume} {70}},\ \bibinfo {pages} {143}
  (\bibinfo {year} {2019})}\BibitemShut {NoStop}%
\bibitem [{\citenamefont {Hollingsworth}\ and\ \citenamefont
  {Dror}(2018)}]{Hollingsworth2018}%
  \BibitemOpen
  \bibfield  {author} {\bibinfo {author} {\bibfnamefont {S.~A.}\ \bibnamefont
  {Hollingsworth}}\ and\ \bibinfo {author} {\bibfnamefont {R.~O.}\ \bibnamefont
  {Dror}},\ }\bibfield  {title} {\bibinfo {title} {Molecular dynamics
  simulation for all},\ }\href@noop {} {\bibfield  {journal} {\bibinfo
  {journal} {Neuron}\ }\textbf {\bibinfo {volume} {99}},\ \bibinfo {pages}
  {1129} (\bibinfo {year} {2018})}\BibitemShut {NoStop}%
\bibitem [{\citenamefont {Ganesan}\ \emph {et~al.}(2017)\citenamefont
  {Ganesan}, \citenamefont {Coote},\ and\ \citenamefont
  {Barakat}}]{Ganesan2017}%
  \BibitemOpen
  \bibfield  {author} {\bibinfo {author} {\bibfnamefont {A.}~\bibnamefont
  {Ganesan}}, \bibinfo {author} {\bibfnamefont {M.~L.}\ \bibnamefont {Coote}},\
  and\ \bibinfo {author} {\bibfnamefont {K.}~\bibnamefont {Barakat}},\
  }\bibfield  {title} {\bibinfo {title} {Molecular dynamics-driven drug
  discovery: leaping forward with confidence},\ }\href@noop {} {\bibfield
  {journal} {\bibinfo  {journal} {Drug discovery today}\ }\textbf {\bibinfo
  {volume} {22}},\ \bibinfo {pages} {249} (\bibinfo {year} {2017})}\BibitemShut
  {NoStop}%
\bibitem [{\citenamefont {Lindorff-Larsen}\ \emph {et~al.}(2011)\citenamefont
  {Lindorff-Larsen}, \citenamefont {Piana}, \citenamefont {Dror},\ and\
  \citenamefont {Shaw}}]{Lindorff-Larsen2011}%
  \BibitemOpen
  \bibfield  {author} {\bibinfo {author} {\bibfnamefont {K.}~\bibnamefont
  {Lindorff-Larsen}}, \bibinfo {author} {\bibfnamefont {S.}~\bibnamefont
  {Piana}}, \bibinfo {author} {\bibfnamefont {R.~O.}\ \bibnamefont {Dror}},\
  and\ \bibinfo {author} {\bibfnamefont {D.~E.}\ \bibnamefont {Shaw}},\
  }\bibfield  {title} {\bibinfo {title} {How fast-folding proteins fold},\
  }\href@noop {} {\bibfield  {journal} {\bibinfo  {journal} {Science}\ }\textbf
  {\bibinfo {volume} {334}},\ \bibinfo {pages} {517} (\bibinfo {year}
  {2011})}\BibitemShut {NoStop}%
\bibitem [{\citenamefont {Piana}\ \emph {et~al.}(2013)\citenamefont {Piana},
  \citenamefont {Lindorff-Larsen},\ and\ \citenamefont {Shaw}}]{Piana2013}%
  \BibitemOpen
  \bibfield  {author} {\bibinfo {author} {\bibfnamefont {S.}~\bibnamefont
  {Piana}}, \bibinfo {author} {\bibfnamefont {K.}~\bibnamefont
  {Lindorff-Larsen}},\ and\ \bibinfo {author} {\bibfnamefont {D.~E.}\
  \bibnamefont {Shaw}},\ }\bibfield  {title} {\bibinfo {title} {Atomic-level
  description of ubiquitin folding},\ }\href@noop {} {\bibfield  {journal}
  {\bibinfo  {journal} {Proceedings of the National Academy of Sciences}\
  }\textbf {\bibinfo {volume} {110}},\ \bibinfo {pages} {5915} (\bibinfo {year}
  {2013})}\BibitemShut {NoStop}%
\bibitem [{\citenamefont {Chodera}\ and\ \citenamefont
  {No{\'e}}(2014)}]{Chodera2014}%
  \BibitemOpen
  \bibfield  {author} {\bibinfo {author} {\bibfnamefont {J.~D.}\ \bibnamefont
  {Chodera}}\ and\ \bibinfo {author} {\bibfnamefont {F.}~\bibnamefont
  {No{\'e}}},\ }\bibfield  {title} {\bibinfo {title} {Markov state models of
  biomolecular conformational dynamics},\ }\href@noop {} {\bibfield  {journal}
  {\bibinfo  {journal} {Current opinion in structural biology}\ }\textbf
  {\bibinfo {volume} {25}},\ \bibinfo {pages} {135} (\bibinfo {year}
  {2014})}\BibitemShut {NoStop}%
\bibitem [{\citenamefont {Plattner}\ \emph {et~al.}(2017)\citenamefont
  {Plattner}, \citenamefont {Doerr}, \citenamefont {De~Fabritiis},\ and\
  \citenamefont {No{\'e}}}]{Plattner2017}%
  \BibitemOpen
  \bibfield  {author} {\bibinfo {author} {\bibfnamefont {N.}~\bibnamefont
  {Plattner}}, \bibinfo {author} {\bibfnamefont {S.}~\bibnamefont {Doerr}},
  \bibinfo {author} {\bibfnamefont {G.}~\bibnamefont {De~Fabritiis}},\ and\
  \bibinfo {author} {\bibfnamefont {F.}~\bibnamefont {No{\'e}}},\ }\bibfield
  {title} {\bibinfo {title} {Complete protein--protein association kinetics in
  atomic detail revealed by molecular dynamics simulations and markov
  modelling},\ }\href@noop {} {\bibfield  {journal} {\bibinfo  {journal}
  {Nature chemistry}\ }\textbf {\bibinfo {volume} {9}},\ \bibinfo {pages}
  {1005} (\bibinfo {year} {2017})}\BibitemShut {NoStop}%
\bibitem [{\citenamefont {Tang}\ \emph {et~al.}(2020)\citenamefont {Tang},
  \citenamefont {Chen},\ and\ \citenamefont {Chang}}]{Tang2020a}%
  \BibitemOpen
  \bibfield  {author} {\bibinfo {author} {\bibfnamefont {Z.}~\bibnamefont
  {Tang}}, \bibinfo {author} {\bibfnamefont {S.-H.}\ \bibnamefont {Chen}},\
  and\ \bibinfo {author} {\bibfnamefont {C.-E.~A.}\ \bibnamefont {Chang}},\
  }\bibfield  {title} {\bibinfo {title} {Transient states and barriers from
  molecular simulations and the milestoning theory: Kinetics in ligand--protein
  recognition and compound design},\ }\href@noop {} {\bibfield  {journal}
  {\bibinfo  {journal} {Journal of chemical theory and computation}\ }\textbf
  {\bibinfo {volume} {16}},\ \bibinfo {pages} {1882} (\bibinfo {year}
  {2020})}\BibitemShut {NoStop}%
\bibitem [{\citenamefont {Wolf}\ \emph {et~al.}(2020)\citenamefont {Wolf},
  \citenamefont {Lickert}, \citenamefont {Bray},\ and\ \citenamefont
  {Stock}}]{Wolf2020}%
  \BibitemOpen
  \bibfield  {author} {\bibinfo {author} {\bibfnamefont {S.}~\bibnamefont
  {Wolf}}, \bibinfo {author} {\bibfnamefont {B.}~\bibnamefont {Lickert}},
  \bibinfo {author} {\bibfnamefont {S.}~\bibnamefont {Bray}},\ and\ \bibinfo
  {author} {\bibfnamefont {G.}~\bibnamefont {Stock}},\ }\bibfield  {title}
  {\bibinfo {title} {Multisecond ligand dissociation dynamics from atomistic
  simulations},\ }\href@noop {} {\bibfield  {journal} {\bibinfo  {journal}
  {Nature communications}\ }\textbf {\bibinfo {volume} {11}},\ \bibinfo {pages}
  {1} (\bibinfo {year} {2020})}\BibitemShut {NoStop}%
\bibitem [{\citenamefont {De~Vivo}\ \emph {et~al.}(2016)\citenamefont
  {De~Vivo}, \citenamefont {Masetti}, \citenamefont {Bottegoni},\ and\
  \citenamefont {Cavalli}}]{DeVivo2016b}%
  \BibitemOpen
  \bibfield  {author} {\bibinfo {author} {\bibfnamefont {M.}~\bibnamefont
  {De~Vivo}}, \bibinfo {author} {\bibfnamefont {M.}~\bibnamefont {Masetti}},
  \bibinfo {author} {\bibfnamefont {G.}~\bibnamefont {Bottegoni}},\ and\
  \bibinfo {author} {\bibfnamefont {A.}~\bibnamefont {Cavalli}},\ }\bibfield
  {title} {\bibinfo {title} {Role of molecular dynamics and related methods in
  drug discovery},\ }\href@noop {} {\bibfield  {journal} {\bibinfo  {journal}
  {Journal of medicinal chemistry}\ }\textbf {\bibinfo {volume} {59}},\
  \bibinfo {pages} {4035} (\bibinfo {year} {2016})}\BibitemShut {NoStop}%
\bibitem [{\citenamefont {Dickson}\ \emph {et~al.}(2017)\citenamefont
  {Dickson}, \citenamefont {Tiwary},\ and\ \citenamefont
  {Vashisth}}]{Dickson2017}%
  \BibitemOpen
  \bibfield  {author} {\bibinfo {author} {\bibfnamefont {A.}~\bibnamefont
  {Dickson}}, \bibinfo {author} {\bibfnamefont {P.}~\bibnamefont {Tiwary}},\
  and\ \bibinfo {author} {\bibfnamefont {H.}~\bibnamefont {Vashisth}},\
  }\bibfield  {title} {\bibinfo {title} {Kinetics of ligand binding through
  advanced computational approaches: a review},\ }\href@noop {} {\bibfield
  {journal} {\bibinfo  {journal} {Current topics in medicinal chemistry}\
  }\textbf {\bibinfo {volume} {17}},\ \bibinfo {pages} {2626} (\bibinfo {year}
  {2017})}\BibitemShut {NoStop}%
\bibitem [{\citenamefont {Ribeiro}\ \emph {et~al.}(2018)\citenamefont
  {Ribeiro}, \citenamefont {Tsai}, \citenamefont {Pramanik}, \citenamefont
  {Wang},\ and\ \citenamefont {Tiwary}}]{Ribeiro2019a}%
  \BibitemOpen
  \bibfield  {author} {\bibinfo {author} {\bibfnamefont {J.~M.~L.}\
  \bibnamefont {Ribeiro}}, \bibinfo {author} {\bibfnamefont {S.-T.}\
  \bibnamefont {Tsai}}, \bibinfo {author} {\bibfnamefont {D.}~\bibnamefont
  {Pramanik}}, \bibinfo {author} {\bibfnamefont {Y.}~\bibnamefont {Wang}},\
  and\ \bibinfo {author} {\bibfnamefont {P.}~\bibnamefont {Tiwary}},\
  }\bibfield  {title} {\bibinfo {title} {Kinetics of ligand--protein
  dissociation from all-atom simulations: Are we there yet?},\ }\href@noop {}
  {\bibfield  {journal} {\bibinfo  {journal} {Biochemistry}\ }\textbf {\bibinfo
  {volume} {58}},\ \bibinfo {pages} {156} (\bibinfo {year} {2018})}\BibitemShut
  {NoStop}%
\bibitem [{\citenamefont {Pietrucci}(2017)}]{Pietrucci2017}%
  \BibitemOpen
  \bibfield  {author} {\bibinfo {author} {\bibfnamefont {F.}~\bibnamefont
  {Pietrucci}},\ }\bibfield  {title} {\bibinfo {title} {Strategies for the
  exploration of free energy landscapes: Unity in diversity and challenges
  ahead},\ }\href@noop {} {\bibfield  {journal} {\bibinfo  {journal} {Reviews
  in Physics}\ }\textbf {\bibinfo {volume} {2}},\ \bibinfo {pages} {32}
  (\bibinfo {year} {2017})}\BibitemShut {NoStop}%
\bibitem [{\citenamefont {Bernardi}\ \emph {et~al.}(2015)\citenamefont
  {Bernardi}, \citenamefont {Melo},\ and\ \citenamefont
  {Schulten}}]{Bernardi2015b}%
  \BibitemOpen
  \bibfield  {author} {\bibinfo {author} {\bibfnamefont {R.~C.}\ \bibnamefont
  {Bernardi}}, \bibinfo {author} {\bibfnamefont {M.~C.}\ \bibnamefont {Melo}},\
  and\ \bibinfo {author} {\bibfnamefont {K.}~\bibnamefont {Schulten}},\
  }\bibfield  {title} {\bibinfo {title} {Enhanced sampling techniques in
  molecular dynamics simulations of biological systems},\ }\href@noop {}
  {\bibfield  {journal} {\bibinfo  {journal} {Biochimica et Biophysica Acta
  (BBA)-General Subjects}\ }\textbf {\bibinfo {volume} {1850}},\ \bibinfo
  {pages} {872} (\bibinfo {year} {2015})}\BibitemShut {NoStop}%
\bibitem [{\citenamefont {Laio}\ and\ \citenamefont
  {Parrinello}(2002)}]{Laio2002}%
  \BibitemOpen
  \bibfield  {author} {\bibinfo {author} {\bibfnamefont {A.}~\bibnamefont
  {Laio}}\ and\ \bibinfo {author} {\bibfnamefont {M.}~\bibnamefont
  {Parrinello}},\ }\bibfield  {title} {\bibinfo {title} {Escaping free-energy
  minima},\ }\href@noop {} {\bibfield  {journal} {\bibinfo  {journal}
  {Proceedings of the National Academy of Sciences}\ }\textbf {\bibinfo
  {volume} {99}},\ \bibinfo {pages} {12562} (\bibinfo {year}
  {2002})}\BibitemShut {NoStop}%
\bibitem [{\citenamefont {Aci-S{\`e}che}\ \emph {et~al.}(2016)\citenamefont
  {Aci-S{\`e}che}, \citenamefont {Ziada}, \citenamefont {Braka}, \citenamefont
  {Arora},\ and\ \citenamefont {Bonnet}}]{Aci-Seche2016}%
  \BibitemOpen
  \bibfield  {author} {\bibinfo {author} {\bibfnamefont {S.}~\bibnamefont
  {Aci-S{\`e}che}}, \bibinfo {author} {\bibfnamefont {S.}~\bibnamefont
  {Ziada}}, \bibinfo {author} {\bibfnamefont {A.}~\bibnamefont {Braka}},
  \bibinfo {author} {\bibfnamefont {R.}~\bibnamefont {Arora}},\ and\ \bibinfo
  {author} {\bibfnamefont {P.}~\bibnamefont {Bonnet}},\ }\bibfield  {title}
  {\bibinfo {title} {Advanced molecular dynamics simulation methods for kinase
  drug discovery},\ }\href@noop {} {\bibfield  {journal} {\bibinfo  {journal}
  {Future medicinal chemistry}\ }\textbf {\bibinfo {volume} {8}},\ \bibinfo
  {pages} {545} (\bibinfo {year} {2016})}\BibitemShut {NoStop}%
\bibitem [{\citenamefont {Cavalli}\ \emph {et~al.}(2015)\citenamefont
  {Cavalli}, \citenamefont {Spitaleri}, \citenamefont {Saladino},\ and\
  \citenamefont {Gervasio}}]{Cavalli2015}%
  \BibitemOpen
  \bibfield  {author} {\bibinfo {author} {\bibfnamefont {A.}~\bibnamefont
  {Cavalli}}, \bibinfo {author} {\bibfnamefont {A.}~\bibnamefont {Spitaleri}},
  \bibinfo {author} {\bibfnamefont {G.}~\bibnamefont {Saladino}},\ and\
  \bibinfo {author} {\bibfnamefont {F.~L.}\ \bibnamefont {Gervasio}},\
  }\bibfield  {title} {\bibinfo {title} {Investigating drug--target association
  and dissociation mechanisms using metadynamics-based algorithms},\
  }\href@noop {} {\bibfield  {journal} {\bibinfo  {journal} {Accounts of
  chemical research}\ }\textbf {\bibinfo {volume} {48}},\ \bibinfo {pages}
  {277} (\bibinfo {year} {2015})}\BibitemShut {NoStop}%
\bibitem [{\citenamefont {Bussi}\ and\ \citenamefont {Laio}(2020)}]{bussi2020}%
  \BibitemOpen
  \bibfield  {author} {\bibinfo {author} {\bibfnamefont {G.}~\bibnamefont
  {Bussi}}\ and\ \bibinfo {author} {\bibfnamefont {A.}~\bibnamefont {Laio}},\
  }\bibfield  {title} {\bibinfo {title} {Using metadynamics to explore complex
  free-energy landscapes},\ }\href@noop {} {\bibfield  {journal} {\bibinfo
  {journal} {Nature Reviews Physics}\ }\textbf {\bibinfo {volume} {2}},\
  \bibinfo {pages} {200} (\bibinfo {year} {2020})}\BibitemShut {NoStop}%
\bibitem [{\citenamefont {Camilloni}\ and\ \citenamefont
  {Pietrucci}(2018)}]{camilloni2018}%
  \BibitemOpen
  \bibfield  {author} {\bibinfo {author} {\bibfnamefont {C.}~\bibnamefont
  {Camilloni}}\ and\ \bibinfo {author} {\bibfnamefont {F.}~\bibnamefont
  {Pietrucci}},\ }\bibfield  {title} {\bibinfo {title} {Advanced simulation
  techniques for the thermodynamic and kinetic characterization of biological
  systems},\ }\href@noop {} {\bibfield  {journal} {\bibinfo  {journal}
  {Advances in Physics: X}\ }\textbf {\bibinfo {volume} {3}},\ \bibinfo {pages}
  {1477531} (\bibinfo {year} {2018})}\BibitemShut {NoStop}%
\bibitem [{\citenamefont {Kokh}\ \emph {et~al.}(2018)\citenamefont {Kokh},
  \citenamefont {Amaral}, \citenamefont {Bomke}, \citenamefont {Gr\"{a}dler},
  \citenamefont {Musil}, \citenamefont {Buchstaller}, \citenamefont {Dreyer},
  \citenamefont {Frech}, \citenamefont {Lowinski}, \citenamefont {Vallee} \emph
  {et~al.}}]{Kokh2018}%
  \BibitemOpen
  \bibfield  {author} {\bibinfo {author} {\bibfnamefont {D.~B.}\ \bibnamefont
  {Kokh}}, \bibinfo {author} {\bibfnamefont {M.}~\bibnamefont {Amaral}},
  \bibinfo {author} {\bibfnamefont {J.}~\bibnamefont {Bomke}}, \bibinfo
  {author} {\bibfnamefont {U.}~\bibnamefont {Gr\"{a}dler}}, \bibinfo {author}
  {\bibfnamefont {D.}~\bibnamefont {Musil}}, \bibinfo {author} {\bibfnamefont
  {H.-P.}\ \bibnamefont {Buchstaller}}, \bibinfo {author} {\bibfnamefont
  {M.~K.}\ \bibnamefont {Dreyer}}, \bibinfo {author} {\bibfnamefont
  {M.}~\bibnamefont {Frech}}, \bibinfo {author} {\bibfnamefont
  {M.}~\bibnamefont {Lowinski}}, \bibinfo {author} {\bibfnamefont
  {F.}~\bibnamefont {Vallee}}, \emph {et~al.},\ }\bibfield  {title} {\bibinfo
  {title} {Estimation of drug-target residence times by $\tau$-random
  acceleration molecular dynamics simulations},\ }\href@noop {} {\bibfield
  {journal} {\bibinfo  {journal} {Journal of chemical theory and computation}\
  }\textbf {\bibinfo {volume} {14}},\ \bibinfo {pages} {3859} (\bibinfo {year}
  {2018})}\BibitemShut {NoStop}%
\bibitem [{\citenamefont {Nunes-Alves}\ \emph {et~al.}(2020)\citenamefont
  {Nunes-Alves}, \citenamefont {Kokh},\ and\ \citenamefont {Wade}}]{Nunes2020}%
  \BibitemOpen
  \bibfield  {author} {\bibinfo {author} {\bibfnamefont {A.}~\bibnamefont
  {Nunes-Alves}}, \bibinfo {author} {\bibfnamefont {D.~B.}\ \bibnamefont
  {Kokh}},\ and\ \bibinfo {author} {\bibfnamefont {R.~C.}\ \bibnamefont
  {Wade}},\ }\bibfield  {title} {\bibinfo {title} {Recent progress in molecular
  simulation methods for drug binding kinetics},\ }\href@noop {} {\bibfield
  {journal} {\bibinfo  {journal} {Current Opinion in Structural Biology}\
  }\textbf {\bibinfo {volume} {64}},\ \bibinfo {pages} {126} (\bibinfo {year}
  {2020})}\BibitemShut {NoStop}%
\bibitem [{\citenamefont {Hummer}(2005)}]{Hummer2005}%
  \BibitemOpen
  \bibfield  {author} {\bibinfo {author} {\bibfnamefont {G.}~\bibnamefont
  {Hummer}},\ }\bibfield  {title} {\bibinfo {title} {Position-dependent
  diffusion coefficients and free energies from bayesian analysis of
  equilibrium and replica molecular dynamics simulations},\ }\href@noop {}
  {\bibfield  {journal} {\bibinfo  {journal} {New Journal of Physics}\ }\textbf
  {\bibinfo {volume} {7}},\ \bibinfo {pages} {34} (\bibinfo {year}
  {2005})}\BibitemShut {NoStop}%
\bibitem [{\citenamefont {Marinelli}\ \emph {et~al.}(2009)\citenamefont
  {Marinelli}, \citenamefont {Pietrucci}, \citenamefont {Laio},\ and\
  \citenamefont {Piana}}]{Marinelli2009}%
  \BibitemOpen
  \bibfield  {author} {\bibinfo {author} {\bibfnamefont {F.}~\bibnamefont
  {Marinelli}}, \bibinfo {author} {\bibfnamefont {F.}~\bibnamefont
  {Pietrucci}}, \bibinfo {author} {\bibfnamefont {A.}~\bibnamefont {Laio}},\
  and\ \bibinfo {author} {\bibfnamefont {S.}~\bibnamefont {Piana}},\ }\bibfield
   {title} {\bibinfo {title} {A kinetic model of trp-cage folding from multiple
  biased molecular dynamics simulations},\ }\href@noop {} {\bibfield  {journal}
  {\bibinfo  {journal} {PLoS computational biology}\ }\textbf {\bibinfo
  {volume} {5}},\ \bibinfo {pages} {e1000452} (\bibinfo {year}
  {2009})}\BibitemShut {NoStop}%
\bibitem [{\citenamefont {Tiwary}\ and\ \citenamefont
  {Parrinello}(2013)}]{Tiwary2013}%
  \BibitemOpen
  \bibfield  {author} {\bibinfo {author} {\bibfnamefont {P.}~\bibnamefont
  {Tiwary}}\ and\ \bibinfo {author} {\bibfnamefont {M.}~\bibnamefont
  {Parrinello}},\ }\bibfield  {title} {\bibinfo {title} {From metadynamics to
  dynamics},\ }\href@noop {} {\bibfield  {journal} {\bibinfo  {journal}
  {Physical review letters}\ }\textbf {\bibinfo {volume} {111}},\ \bibinfo
  {pages} {230602} (\bibinfo {year} {2013})}\BibitemShut {NoStop}%
\bibitem [{\citenamefont {Stelzl}\ and\ \citenamefont
  {Hummer}(2017)}]{Stelzl2017a}%
  \BibitemOpen
  \bibfield  {author} {\bibinfo {author} {\bibfnamefont {L.~S.}\ \bibnamefont
  {Stelzl}}\ and\ \bibinfo {author} {\bibfnamefont {G.}~\bibnamefont
  {Hummer}},\ }\bibfield  {title} {\bibinfo {title} {Kinetics from replica
  exchange molecular dynamics simulations},\ }\href@noop {} {\bibfield
  {journal} {\bibinfo  {journal} {Journal of chemical theory and computation}\
  }\textbf {\bibinfo {volume} {13}},\ \bibinfo {pages} {3927} (\bibinfo {year}
  {2017})}\BibitemShut {NoStop}%
\bibitem [{\citenamefont {Stelzl}\ \emph {et~al.}(2017)\citenamefont {Stelzl},
  \citenamefont {Kells}, \citenamefont {Rosta},\ and\ \citenamefont
  {Hummer}}]{Stelzl2017}%
  \BibitemOpen
  \bibfield  {author} {\bibinfo {author} {\bibfnamefont {L.~S.}\ \bibnamefont
  {Stelzl}}, \bibinfo {author} {\bibfnamefont {A.}~\bibnamefont {Kells}},
  \bibinfo {author} {\bibfnamefont {E.}~\bibnamefont {Rosta}},\ and\ \bibinfo
  {author} {\bibfnamefont {G.}~\bibnamefont {Hummer}},\ }\bibfield  {title}
  {\bibinfo {title} {Dynamic histogram analysis to determine free energies and
  rates from biased simulations},\ }\href@noop {} {\bibfield  {journal}
  {\bibinfo  {journal} {Journal of chemical theory and computation}\ }\textbf
  {\bibinfo {volume} {13}},\ \bibinfo {pages} {6328} (\bibinfo {year}
  {2017})}\BibitemShut {NoStop}%
\bibitem [{\citenamefont {Donati}\ and\ \citenamefont
  {Keller}(2018)}]{Donati2018}%
  \BibitemOpen
  \bibfield  {author} {\bibinfo {author} {\bibfnamefont {L.}~\bibnamefont
  {Donati}}\ and\ \bibinfo {author} {\bibfnamefont {B.~G.}\ \bibnamefont
  {Keller}},\ }\bibfield  {title} {\bibinfo {title} {Girsanov reweighting for
  metadynamics simulations},\ }\href@noop {} {\bibfield  {journal} {\bibinfo
  {journal} {The Journal of chemical physics}\ }\textbf {\bibinfo {volume}
  {149}},\ \bibinfo {pages} {072335} (\bibinfo {year} {2018})}\BibitemShut
  {NoStop}%
\bibitem [{\citenamefont {Sch\"afer}\ and\ \citenamefont
  {Settanni}(2020)}]{Schafer2020b}%
  \BibitemOpen
  \bibfield  {author} {\bibinfo {author} {\bibfnamefont {T.~M.}\ \bibnamefont
  {Sch\"afer}}\ and\ \bibinfo {author} {\bibfnamefont {G.}~\bibnamefont
  {Settanni}},\ }\bibfield  {title} {\bibinfo {title} {Data reweighting in
  metadynamics simulations},\ }\href@noop {} {\bibfield  {journal} {\bibinfo
  {journal} {Journal of chemical theory and computation}\ }\textbf {\bibinfo
  {volume} {16}},\ \bibinfo {pages} {2042} (\bibinfo {year}
  {2020})}\BibitemShut {NoStop}%
\bibitem [{\citenamefont {Kieninger}\ \emph {et~al.}(2020)\citenamefont
  {Kieninger}, \citenamefont {Donati},\ and\ \citenamefont
  {Keller}}]{Kieninger2020}%
  \BibitemOpen
  \bibfield  {author} {\bibinfo {author} {\bibfnamefont {S.}~\bibnamefont
  {Kieninger}}, \bibinfo {author} {\bibfnamefont {L.}~\bibnamefont {Donati}},\
  and\ \bibinfo {author} {\bibfnamefont {B.~G.}\ \bibnamefont {Keller}},\
  }\bibfield  {title} {\bibinfo {title} {Dynamical reweighting methods for
  markov models},\ }\href@noop {} {\bibfield  {journal} {\bibinfo  {journal}
  {Current opinion in structural biology}\ }\textbf {\bibinfo {volume} {61}},\
  \bibinfo {pages} {124} (\bibinfo {year} {2020})}\BibitemShut {NoStop}%
\bibitem [{\citenamefont {Linker}\ \emph {et~al.}(2020)\citenamefont {Linker},
  \citenamefont {Wei{\ss}},\ and\ \citenamefont {Riniker}}]{Linker2020}%
  \BibitemOpen
  \bibfield  {author} {\bibinfo {author} {\bibfnamefont {S.~M.}\ \bibnamefont
  {Linker}}, \bibinfo {author} {\bibfnamefont {R.~G.}\ \bibnamefont
  {Wei{\ss}}},\ and\ \bibinfo {author} {\bibfnamefont {S.}~\bibnamefont
  {Riniker}},\ }\bibfield  {title} {\bibinfo {title} {Connecting dynamic
  reweighting algorithms: Derivation of the dynamic reweighting family tree},\
  }\href@noop {} {\bibfield  {journal} {\bibinfo  {journal} {The Journal of
  Chemical Physics}\ }\textbf {\bibinfo {volume} {153}},\ \bibinfo {pages}
  {234106} (\bibinfo {year} {2020})}\BibitemShut {NoStop}%
\bibitem [{\citenamefont {Tiwary}\ \emph {et~al.}(2015)\citenamefont {Tiwary},
  \citenamefont {Limongelli}, \citenamefont {Salvalaglio},\ and\ \citenamefont
  {Parrinello}}]{Tiwary2015a}%
  \BibitemOpen
  \bibfield  {author} {\bibinfo {author} {\bibfnamefont {P.}~\bibnamefont
  {Tiwary}}, \bibinfo {author} {\bibfnamefont {V.}~\bibnamefont {Limongelli}},
  \bibinfo {author} {\bibfnamefont {M.}~\bibnamefont {Salvalaglio}},\ and\
  \bibinfo {author} {\bibfnamefont {M.}~\bibnamefont {Parrinello}},\ }\bibfield
   {title} {\bibinfo {title} {Kinetics of protein--ligand unbinding: Predicting
  pathways, rates, and rate-limiting steps},\ }\href@noop {} {\bibfield
  {journal} {\bibinfo  {journal} {Proceedings of the National Academy of
  Sciences}\ }\textbf {\bibinfo {volume} {112}},\ \bibinfo {pages} {E386}
  (\bibinfo {year} {2015})}\BibitemShut {NoStop}%
\bibitem [{\citenamefont {Tiwary}\ \emph {et~al.}(2017)\citenamefont {Tiwary},
  \citenamefont {Mondal},\ and\ \citenamefont {Berne}}]{Tiwary2017}%
  \BibitemOpen
  \bibfield  {author} {\bibinfo {author} {\bibfnamefont {P.}~\bibnamefont
  {Tiwary}}, \bibinfo {author} {\bibfnamefont {J.}~\bibnamefont {Mondal}},\
  and\ \bibinfo {author} {\bibfnamefont {B.~J.}\ \bibnamefont {Berne}},\
  }\bibfield  {title} {\bibinfo {title} {How and when does an anticancer drug
  leave its binding site?},\ }\href@noop {} {\bibfield  {journal} {\bibinfo
  {journal} {Science advances}\ }\textbf {\bibinfo {volume} {3}},\ \bibinfo
  {pages} {e1700014} (\bibinfo {year} {2017})}\BibitemShut {NoStop}%
\bibitem [{\citenamefont {Casasnovas}\ \emph {et~al.}(2017)\citenamefont
  {Casasnovas}, \citenamefont {Limongelli}, \citenamefont {Tiwary},
  \citenamefont {Carloni},\ and\ \citenamefont {Parrinello}}]{Casasnovas2017}%
  \BibitemOpen
  \bibfield  {author} {\bibinfo {author} {\bibfnamefont {R.}~\bibnamefont
  {Casasnovas}}, \bibinfo {author} {\bibfnamefont {V.}~\bibnamefont
  {Limongelli}}, \bibinfo {author} {\bibfnamefont {P.}~\bibnamefont {Tiwary}},
  \bibinfo {author} {\bibfnamefont {P.}~\bibnamefont {Carloni}},\ and\ \bibinfo
  {author} {\bibfnamefont {M.}~\bibnamefont {Parrinello}},\ }\bibfield  {title}
  {\bibinfo {title} {Unbinding kinetics of a p38 map kinase type ii inhibitor
  from metadynamics simulations},\ }\href@noop {} {\bibfield  {journal}
  {\bibinfo  {journal} {Journal of the American Chemical Society}\ }\textbf
  {\bibinfo {volume} {139}},\ \bibinfo {pages} {4780} (\bibinfo {year}
  {2017})}\BibitemShut {NoStop}%
\bibitem [{\citenamefont {Sun}\ \emph {et~al.}(2017)\citenamefont {Sun},
  \citenamefont {Li}, \citenamefont {Shen}, \citenamefont {Li}, \citenamefont
  {Kang},\ and\ \citenamefont {Hou}}]{Sun2017a}%
  \BibitemOpen
  \bibfield  {author} {\bibinfo {author} {\bibfnamefont {H.}~\bibnamefont
  {Sun}}, \bibinfo {author} {\bibfnamefont {Y.}~\bibnamefont {Li}}, \bibinfo
  {author} {\bibfnamefont {M.}~\bibnamefont {Shen}}, \bibinfo {author}
  {\bibfnamefont {D.}~\bibnamefont {Li}}, \bibinfo {author} {\bibfnamefont
  {Y.}~\bibnamefont {Kang}},\ and\ \bibinfo {author} {\bibfnamefont
  {T.}~\bibnamefont {Hou}},\ }\bibfield  {title} {\bibinfo {title}
  {Characterizing drug--target residence time with metadynamics: How to achieve
  dissociation rate efficiently without losing accuracy against time-consuming
  approaches},\ }\href@noop {} {\bibfield  {journal} {\bibinfo  {journal}
  {Journal of chemical information and modeling}\ }\textbf {\bibinfo {volume}
  {57}},\ \bibinfo {pages} {1895} (\bibinfo {year} {2017})}\BibitemShut
  {NoStop}%
\bibitem [{\citenamefont {Pramanik}\ \emph {et~al.}(2019)\citenamefont
  {Pramanik}, \citenamefont {Smith}, \citenamefont {Kells},\ and\ \citenamefont
  {Tiwary}}]{Pramanik2019}%
  \BibitemOpen
  \bibfield  {author} {\bibinfo {author} {\bibfnamefont {D.}~\bibnamefont
  {Pramanik}}, \bibinfo {author} {\bibfnamefont {Z.}~\bibnamefont {Smith}},
  \bibinfo {author} {\bibfnamefont {A.}~\bibnamefont {Kells}},\ and\ \bibinfo
  {author} {\bibfnamefont {P.}~\bibnamefont {Tiwary}},\ }\bibfield  {title}
  {\bibinfo {title} {Can one trust kinetic and thermodynamic observables from
  biased metadynamics simulations?: Detailed quantitative benchmarks on
  millimolar drug fragment dissociation},\ }\href@noop {} {\bibfield  {journal}
  {\bibinfo  {journal} {The Journal of Physical Chemistry B}\ }\textbf
  {\bibinfo {volume} {123}},\ \bibinfo {pages} {3672} (\bibinfo {year}
  {2019})}\BibitemShut {NoStop}%
\bibitem [{\citenamefont {Zou}\ \emph {et~al.}(2020)\citenamefont {Zou},
  \citenamefont {Zhou}, \citenamefont {Wang}, \citenamefont {Kuang},
  \citenamefont {{\AA}gren}, \citenamefont {Wu},\ and\ \citenamefont
  {Tu}}]{Zou2020}%
  \BibitemOpen
  \bibfield  {author} {\bibinfo {author} {\bibfnamefont {R.}~\bibnamefont
  {Zou}}, \bibinfo {author} {\bibfnamefont {Y.}~\bibnamefont {Zhou}}, \bibinfo
  {author} {\bibfnamefont {Y.}~\bibnamefont {Wang}}, \bibinfo {author}
  {\bibfnamefont {G.}~\bibnamefont {Kuang}}, \bibinfo {author} {\bibfnamefont
  {H.}~\bibnamefont {{\AA}gren}}, \bibinfo {author} {\bibfnamefont
  {J.}~\bibnamefont {Wu}},\ and\ \bibinfo {author} {\bibfnamefont
  {Y.}~\bibnamefont {Tu}},\ }\bibfield  {title} {\bibinfo {title} {Free energy
  profile and kinetics of coupled folding and binding of the intrinsically
  disordered protein p53 with mdm2},\ }\href@noop {} {\bibfield  {journal}
  {\bibinfo  {journal} {Journal of chemical information and modeling}\ }\textbf
  {\bibinfo {volume} {60}},\ \bibinfo {pages} {1551} (\bibinfo {year}
  {2020})}\BibitemShut {NoStop}%
\bibitem [{\citenamefont {Lamim~Ribeiro}\ \emph {et~al.}(2020)\citenamefont
  {Lamim~Ribeiro}, \citenamefont {Provasi},\ and\ \citenamefont
  {Filizola}}]{LamimRibeiro2020}%
  \BibitemOpen
  \bibfield  {author} {\bibinfo {author} {\bibfnamefont {J.~M.}\ \bibnamefont
  {Lamim~Ribeiro}}, \bibinfo {author} {\bibfnamefont {D.}~\bibnamefont
  {Provasi}},\ and\ \bibinfo {author} {\bibfnamefont {M.}~\bibnamefont
  {Filizola}},\ }\bibfield  {title} {\bibinfo {title} {A combination of machine
  learning and infrequent metadynamics to efficiently predict kinetic rates,
  transition states, and molecular determinants of drug dissociation from g
  protein-coupled receptors},\ }\href@noop {} {\bibfield  {journal} {\bibinfo
  {journal} {The Journal of Chemical Physics}\ }\textbf {\bibinfo {volume}
  {153}},\ \bibinfo {pages} {124105} (\bibinfo {year} {2020})}\BibitemShut
  {NoStop}%
\bibitem [{\citenamefont {Shekhar}\ \emph {et~al.}(2021)\citenamefont
  {Shekhar}, \citenamefont {Smith}, \citenamefont {Seeliger},\ and\
  \citenamefont {Tiwary}}]{Shekhar2021}%
  \BibitemOpen
  \bibfield  {author} {\bibinfo {author} {\bibfnamefont {M.}~\bibnamefont
  {Shekhar}}, \bibinfo {author} {\bibfnamefont {Z.}~\bibnamefont {Smith}},
  \bibinfo {author} {\bibfnamefont {M.}~\bibnamefont {Seeliger}},\ and\
  \bibinfo {author} {\bibfnamefont {P.}~\bibnamefont {Tiwary}},\ }\bibfield
  {title} {\bibinfo {title} {Protein flexibility and dissociation pathway
  differentiation can explain onset of resistance mutations in kinases},\
  }\href@noop {} {\bibfield  {journal} {\bibinfo  {journal} {bioRxiv}\ }
  (\bibinfo {year} {2021})}\BibitemShut {NoStop}%
\bibitem [{\citenamefont {Grubm{\"u}ller}(1995)}]{Grubmuller1995a}%
  \BibitemOpen
  \bibfield  {author} {\bibinfo {author} {\bibfnamefont {H.}~\bibnamefont
  {Grubm{\"u}ller}},\ }\bibfield  {title} {\bibinfo {title} {Predicting slow
  structural transitions in macromolecular systems: Conformational flooding},\
  }\href@noop {} {\bibfield  {journal} {\bibinfo  {journal} {Physical Review
  E}\ }\textbf {\bibinfo {volume} {52}},\ \bibinfo {pages} {2893} (\bibinfo
  {year} {1995})}\BibitemShut {NoStop}%
\bibitem [{\citenamefont {Voter}(1997)}]{Voter1997a}%
  \BibitemOpen
  \bibfield  {author} {\bibinfo {author} {\bibfnamefont {A.~F.}\ \bibnamefont
  {Voter}},\ }\bibfield  {title} {\bibinfo {title} {Hyperdynamics: Accelerated
  molecular dynamics of infrequent events},\ }\href@noop {} {\bibfield
  {journal} {\bibinfo  {journal} {Physical Review Letters}\ }\textbf {\bibinfo
  {volume} {78}},\ \bibinfo {pages} {3908} (\bibinfo {year}
  {1997})}\BibitemShut {NoStop}%
\bibitem [{\citenamefont {Salvalaglio}\ \emph {et~al.}(2014)\citenamefont
  {Salvalaglio}, \citenamefont {Tiwary},\ and\ \citenamefont
  {Parrinello}}]{Salvalaglio2014}%
  \BibitemOpen
  \bibfield  {author} {\bibinfo {author} {\bibfnamefont {M.}~\bibnamefont
  {Salvalaglio}}, \bibinfo {author} {\bibfnamefont {P.}~\bibnamefont
  {Tiwary}},\ and\ \bibinfo {author} {\bibfnamefont {M.}~\bibnamefont
  {Parrinello}},\ }\bibfield  {title} {\bibinfo {title} {Assessing the
  reliability of the dynamics reconstructed from metadynamics},\ }\href@noop {}
  {\bibfield  {journal} {\bibinfo  {journal} {Journal of chemical theory and
  computation}\ }\textbf {\bibinfo {volume} {10}},\ \bibinfo {pages} {1420}
  (\bibinfo {year} {2014})}\BibitemShut {NoStop}%
\bibitem [{\citenamefont {Dickson}(2018)}]{Dickson2018}%
  \BibitemOpen
  \bibfield  {author} {\bibinfo {author} {\bibfnamefont {B.~M.}\ \bibnamefont
  {Dickson}},\ }\bibfield  {title} {\bibinfo {title} {Erroneous rates and false
  statistical confirmations from infrequent metadynamics and other equivalent
  violations of the hyperdynamics paradigm},\ }\href@noop {} {\bibfield
  {journal} {\bibinfo  {journal} {Journal of chemical theory and computation}\
  }\textbf {\bibinfo {volume} {15}},\ \bibinfo {pages} {78} (\bibinfo {year}
  {2018})}\BibitemShut {NoStop}%
\bibitem [{\citenamefont {Khan}\ \emph {et~al.}(2020)\citenamefont {Khan},
  \citenamefont {Dickson},\ and\ \citenamefont {Peters}}]{Khan2020}%
  \BibitemOpen
  \bibfield  {author} {\bibinfo {author} {\bibfnamefont {S.~A.}\ \bibnamefont
  {Khan}}, \bibinfo {author} {\bibfnamefont {B.~M.}\ \bibnamefont {Dickson}},\
  and\ \bibinfo {author} {\bibfnamefont {B.}~\bibnamefont {Peters}},\
  }\bibfield  {title} {\bibinfo {title} {How fluxional reactants limit the
  accuracy/efficiency of infrequent metadynamics},\ }\href@noop {} {\bibfield
  {journal} {\bibinfo  {journal} {The Journal of Chemical Physics}\ }\textbf
  {\bibinfo {volume} {153}},\ \bibinfo {pages} {054125} (\bibinfo {year}
  {2020})}\BibitemShut {NoStop}%
\bibitem [{\citenamefont {Callegari}\ \emph {et~al.}(2017)\citenamefont
  {Callegari}, \citenamefont {Lodola}, \citenamefont {Pala}, \citenamefont
  {Rivara}, \citenamefont {Mor}, \citenamefont {Rizzi},\ and\ \citenamefont
  {Capelli}}]{Callegari2017}%
  \BibitemOpen
  \bibfield  {author} {\bibinfo {author} {\bibfnamefont {D.}~\bibnamefont
  {Callegari}}, \bibinfo {author} {\bibfnamefont {A.}~\bibnamefont {Lodola}},
  \bibinfo {author} {\bibfnamefont {D.}~\bibnamefont {Pala}}, \bibinfo {author}
  {\bibfnamefont {S.}~\bibnamefont {Rivara}}, \bibinfo {author} {\bibfnamefont
  {M.}~\bibnamefont {Mor}}, \bibinfo {author} {\bibfnamefont {A.}~\bibnamefont
  {Rizzi}},\ and\ \bibinfo {author} {\bibfnamefont {A.~M.}\ \bibnamefont
  {Capelli}},\ }\bibfield  {title} {\bibinfo {title} {Metadynamics simulations
  distinguish short-and long-residence-time inhibitors of cyclin-dependent
  kinase 8},\ }\href@noop {} {\bibfield  {journal} {\bibinfo  {journal}
  {Journal of chemical information and modeling}\ }\textbf {\bibinfo {volume}
  {57}},\ \bibinfo {pages} {159} (\bibinfo {year} {2017})}\BibitemShut
  {NoStop}%
\bibitem [{\citenamefont {Wang}\ \emph {et~al.}(2018)\citenamefont {Wang},
  \citenamefont {Valsson}, \citenamefont {Tiwary}, \citenamefont {Parrinello},\
  and\ \citenamefont {Lindorff-Larsen}}]{Wang2018}%
  \BibitemOpen
  \bibfield  {author} {\bibinfo {author} {\bibfnamefont {Y.}~\bibnamefont
  {Wang}}, \bibinfo {author} {\bibfnamefont {O.}~\bibnamefont {Valsson}},
  \bibinfo {author} {\bibfnamefont {P.}~\bibnamefont {Tiwary}}, \bibinfo
  {author} {\bibfnamefont {M.}~\bibnamefont {Parrinello}},\ and\ \bibinfo
  {author} {\bibfnamefont {K.}~\bibnamefont {Lindorff-Larsen}},\ }\bibfield
  {title} {\bibinfo {title} {Frequency adaptive metadynamics for the
  calculation of rare-event kinetics},\ }\href@noop {} {\bibfield  {journal}
  {\bibinfo  {journal} {The Journal of chemical physics}\ }\textbf {\bibinfo
  {volume} {149}},\ \bibinfo {pages} {072309} (\bibinfo {year}
  {2018})}\BibitemShut {NoStop}%
\bibitem [{\citenamefont {Hummer}\ and\ \citenamefont
  {Szabo}(2003)}]{Hummer2003}%
  \BibitemOpen
  \bibfield  {author} {\bibinfo {author} {\bibfnamefont {G.}~\bibnamefont
  {Hummer}}\ and\ \bibinfo {author} {\bibfnamefont {A.}~\bibnamefont {Szabo}},\
  }\bibfield  {title} {\bibinfo {title} {Kinetics from nonequilibrium
  single-molecule pulling experiments},\ }\href@noop {} {\bibfield  {journal}
  {\bibinfo  {journal} {Biophysical journal}\ }\textbf {\bibinfo {volume}
  {85}},\ \bibinfo {pages} {5} (\bibinfo {year} {2003})}\BibitemShut {NoStop}%
\bibitem [{\citenamefont {Dudko}\ \emph {et~al.}(2006)\citenamefont {Dudko},
  \citenamefont {Hummer},\ and\ \citenamefont {Szabo}}]{Dudko2006}%
  \BibitemOpen
  \bibfield  {author} {\bibinfo {author} {\bibfnamefont {O.~K.}\ \bibnamefont
  {Dudko}}, \bibinfo {author} {\bibfnamefont {G.}~\bibnamefont {Hummer}},\ and\
  \bibinfo {author} {\bibfnamefont {A.}~\bibnamefont {Szabo}},\ }\bibfield
  {title} {\bibinfo {title} {Intrinsic rates and activation free energies from
  single-molecule pulling experiments},\ }\href@noop {} {\bibfield  {journal}
  {\bibinfo  {journal} {Physical review letters}\ }\textbf {\bibinfo {volume}
  {96}},\ \bibinfo {pages} {108101} (\bibinfo {year} {2006})}\BibitemShut
  {NoStop}%
\bibitem [{\citenamefont {Cossio}\ \emph {et~al.}(2016)\citenamefont {Cossio},
  \citenamefont {Hummer},\ and\ \citenamefont {Szabo}}]{Cossio2016}%
  \BibitemOpen
  \bibfield  {author} {\bibinfo {author} {\bibfnamefont {P.}~\bibnamefont
  {Cossio}}, \bibinfo {author} {\bibfnamefont {G.}~\bibnamefont {Hummer}},\
  and\ \bibinfo {author} {\bibfnamefont {A.}~\bibnamefont {Szabo}},\ }\bibfield
   {title} {\bibinfo {title} {Kinetic ductility and force-spike resistance of
  proteins from single-molecule force spectroscopy},\ }\href@noop {} {\bibfield
   {journal} {\bibinfo  {journal} {Biophysical journal}\ }\textbf {\bibinfo
  {volume} {111}},\ \bibinfo {pages} {832} (\bibinfo {year}
  {2016})}\BibitemShut {NoStop}%
\bibitem [{\citenamefont {Kramers}(1940)}]{Kramers1940}%
  \BibitemOpen
  \bibfield  {author} {\bibinfo {author} {\bibfnamefont {H.~A.}\ \bibnamefont
  {Kramers}},\ }\bibfield  {title} {\bibinfo {title} {Brownian motion in a
  field of force and the diffusion model of chemical reactions},\ }\href@noop
  {} {\bibfield  {journal} {\bibinfo  {journal} {Physica}\ }\textbf {\bibinfo
  {volume} {7}},\ \bibinfo {pages} {284} (\bibinfo {year} {1940})}\BibitemShut
  {NoStop}%
\bibitem [{\citenamefont {H{\"a}nggi}\ \emph {et~al.}(1990)\citenamefont
  {H{\"a}nggi}, \citenamefont {Talkner},\ and\ \citenamefont
  {Borkovec}}]{Hanggi1990}%
  \BibitemOpen
  \bibfield  {author} {\bibinfo {author} {\bibfnamefont {P.}~\bibnamefont
  {H{\"a}nggi}}, \bibinfo {author} {\bibfnamefont {P.}~\bibnamefont
  {Talkner}},\ and\ \bibinfo {author} {\bibfnamefont {M.}~\bibnamefont
  {Borkovec}},\ }\bibfield  {title} {\bibinfo {title} {Reaction-rate theory:
  fifty years after kramers},\ }\href@noop {} {\bibfield  {journal} {\bibinfo
  {journal} {Reviews of modern physics}\ }\textbf {\bibinfo {volume} {62}},\
  \bibinfo {pages} {251} (\bibinfo {year} {1990})}\BibitemShut {NoStop}%
\bibitem [{\citenamefont {Kontopidis}\ \emph {et~al.}(2006)\citenamefont
  {Kontopidis}, \citenamefont {McInnes}, \citenamefont {Pandalaneni},
  \citenamefont {McNae}, \citenamefont {Gibson}, \citenamefont {Mezna},
  \citenamefont {Thomas}, \citenamefont {Wood}, \citenamefont {Wang},
  \citenamefont {Walkinshaw} \emph {et~al.}}]{Kontopidis2006}%
  \BibitemOpen
  \bibfield  {author} {\bibinfo {author} {\bibfnamefont {G.}~\bibnamefont
  {Kontopidis}}, \bibinfo {author} {\bibfnamefont {C.}~\bibnamefont {McInnes}},
  \bibinfo {author} {\bibfnamefont {S.~R.}\ \bibnamefont {Pandalaneni}},
  \bibinfo {author} {\bibfnamefont {I.}~\bibnamefont {McNae}}, \bibinfo
  {author} {\bibfnamefont {D.}~\bibnamefont {Gibson}}, \bibinfo {author}
  {\bibfnamefont {M.}~\bibnamefont {Mezna}}, \bibinfo {author} {\bibfnamefont
  {M.}~\bibnamefont {Thomas}}, \bibinfo {author} {\bibfnamefont
  {G.}~\bibnamefont {Wood}}, \bibinfo {author} {\bibfnamefont {S.}~\bibnamefont
  {Wang}}, \bibinfo {author} {\bibfnamefont {M.~D.}\ \bibnamefont
  {Walkinshaw}}, \emph {et~al.},\ }\bibfield  {title} {\bibinfo {title}
  {Differential binding of inhibitors to active and inactive cdk2 provides
  insights for drug design},\ }\href@noop {} {\bibfield  {journal} {\bibinfo
  {journal} {Chemistry \& biology}\ }\textbf {\bibinfo {volume} {13}},\
  \bibinfo {pages} {201} (\bibinfo {year} {2006})}\BibitemShut {NoStop}%
\bibitem [{\citenamefont {Shapiro}(2006)}]{Shapiro2006}%
  \BibitemOpen
  \bibfield  {author} {\bibinfo {author} {\bibfnamefont {G.~I.}\ \bibnamefont
  {Shapiro}},\ }\bibfield  {title} {\bibinfo {title} {Cyclin-dependent kinase
  pathways as targets for cancer treatment},\ }\href@noop {} {\bibfield
  {journal} {\bibinfo  {journal} {Journal of clinical oncology}\ }\textbf
  {\bibinfo {volume} {24}},\ \bibinfo {pages} {1770} (\bibinfo {year}
  {2006})}\BibitemShut {NoStop}%
\bibitem [{\citenamefont {Barducci}\ \emph {et~al.}(2008)\citenamefont
  {Barducci}, \citenamefont {Bussi},\ and\ \citenamefont
  {Parrinello}}]{Barducci2008}%
  \BibitemOpen
  \bibfield  {author} {\bibinfo {author} {\bibfnamefont {A.}~\bibnamefont
  {Barducci}}, \bibinfo {author} {\bibfnamefont {G.}~\bibnamefont {Bussi}},\
  and\ \bibinfo {author} {\bibfnamefont {M.}~\bibnamefont {Parrinello}},\
  }\bibfield  {title} {\bibinfo {title} {Well-tempered metadynamics: a smoothly
  converging and tunable free-energy method},\ }\href@noop {} {\bibfield
  {journal} {\bibinfo  {journal} {Physical review letters}\ }\textbf {\bibinfo
  {volume} {100}},\ \bibinfo {pages} {020603} (\bibinfo {year}
  {2008})}\BibitemShut {NoStop}%
\bibitem [{\citenamefont {Dunbar~Jr}\ \emph {et~al.}(2013)\citenamefont
  {Dunbar~Jr}, \citenamefont {Smith}, \citenamefont {Damm-Ganamet},
  \citenamefont {Ahmed}, \citenamefont {Esposito}, \citenamefont {Delproposto},
  \citenamefont {Chinnaswamy}, \citenamefont {Kang}, \citenamefont {Kubish},
  \citenamefont {Gestwicki} \emph {et~al.}}]{Dunbar2013}%
  \BibitemOpen
  \bibfield  {author} {\bibinfo {author} {\bibfnamefont {J.~B.}\ \bibnamefont
  {Dunbar~Jr}}, \bibinfo {author} {\bibfnamefont {R.~D.}\ \bibnamefont
  {Smith}}, \bibinfo {author} {\bibfnamefont {K.~L.}\ \bibnamefont
  {Damm-Ganamet}}, \bibinfo {author} {\bibfnamefont {A.}~\bibnamefont {Ahmed}},
  \bibinfo {author} {\bibfnamefont {E.~X.}\ \bibnamefont {Esposito}}, \bibinfo
  {author} {\bibfnamefont {J.}~\bibnamefont {Delproposto}}, \bibinfo {author}
  {\bibfnamefont {K.}~\bibnamefont {Chinnaswamy}}, \bibinfo {author}
  {\bibfnamefont {Y.-N.}\ \bibnamefont {Kang}}, \bibinfo {author}
  {\bibfnamefont {G.}~\bibnamefont {Kubish}}, \bibinfo {author} {\bibfnamefont
  {J.~E.}\ \bibnamefont {Gestwicki}}, \emph {et~al.},\ }\bibfield  {title}
  {\bibinfo {title} {Csar data set release 2012: ligands, affinities,
  complexes, and docking decoys},\ }\href@noop {} {\bibfield  {journal}
  {\bibinfo  {journal} {Journal of chemical information and modeling}\ }\textbf
  {\bibinfo {volume} {53}},\ \bibinfo {pages} {1842} (\bibinfo {year}
  {2013})}\BibitemShut {NoStop}%
\bibitem [{\citenamefont {Darve}\ \emph {et~al.}(2008)\citenamefont {Darve},
  \citenamefont {Rodr{\'\i}guez-G{\'o}mez},\ and\ \citenamefont
  {Pohorille}}]{Darve2008}%
  \BibitemOpen
  \bibfield  {author} {\bibinfo {author} {\bibfnamefont {E.}~\bibnamefont
  {Darve}}, \bibinfo {author} {\bibfnamefont {D.}~\bibnamefont
  {Rodr{\'\i}guez-G{\'o}mez}},\ and\ \bibinfo {author} {\bibfnamefont
  {A.}~\bibnamefont {Pohorille}},\ }\bibfield  {title} {\bibinfo {title}
  {Adaptive biasing force method for scalar and vector free energy
  calculations},\ }\href@noop {} {\bibfield  {journal} {\bibinfo  {journal}
  {The Journal of chemical physics}\ }\textbf {\bibinfo {volume} {128}},\
  \bibinfo {pages} {144120} (\bibinfo {year} {2008})}\BibitemShut {NoStop}%
\bibitem [{\citenamefont {Henin}\ \emph {et~al.}(2010)\citenamefont {Henin},
  \citenamefont {Fiorin}, \citenamefont {Chipot},\ and\ \citenamefont
  {Klein}}]{Henin2010}%
  \BibitemOpen
  \bibfield  {author} {\bibinfo {author} {\bibfnamefont {J.}~\bibnamefont
  {Henin}}, \bibinfo {author} {\bibfnamefont {G.}~\bibnamefont {Fiorin}},
  \bibinfo {author} {\bibfnamefont {C.}~\bibnamefont {Chipot}},\ and\ \bibinfo
  {author} {\bibfnamefont {M.~L.}\ \bibnamefont {Klein}},\ }\bibfield  {title}
  {\bibinfo {title} {Exploring multidimensional free energy landscapes using
  time-dependent biases on collective variables},\ }\href@noop {} {\bibfield
  {journal} {\bibinfo  {journal} {Journal of chemical theory and computation}\
  }\textbf {\bibinfo {volume} {6}},\ \bibinfo {pages} {35} (\bibinfo {year}
  {2010})}\BibitemShut {NoStop}%
\bibitem [{\citenamefont {Marchi}\ and\ \citenamefont
  {Ballone}(1999)}]{Marchi1999}%
  \BibitemOpen
  \bibfield  {author} {\bibinfo {author} {\bibfnamefont {M.}~\bibnamefont
  {Marchi}}\ and\ \bibinfo {author} {\bibfnamefont {P.}~\bibnamefont
  {Ballone}},\ }\bibfield  {title} {\bibinfo {title} {Adiabatic bias molecular
  dynamics: a method to navigate the conformational space of complex molecular
  systems},\ }\href@noop {} {\bibfield  {journal} {\bibinfo  {journal} {The
  Journal of chemical physics}\ }\textbf {\bibinfo {volume} {110}},\ \bibinfo
  {pages} {3697} (\bibinfo {year} {1999})}\BibitemShut {NoStop}%
\bibitem [{\citenamefont {Paci}\ and\ \citenamefont
  {Karplus}(1999)}]{Paci1999}%
  \BibitemOpen
  \bibfield  {author} {\bibinfo {author} {\bibfnamefont {E.}~\bibnamefont
  {Paci}}\ and\ \bibinfo {author} {\bibfnamefont {M.}~\bibnamefont {Karplus}},\
  }\bibfield  {title} {\bibinfo {title} {Forced unfolding of fibronectin type 3
  modules: an analysis by biased molecular dynamics simulations},\ }\href@noop
  {} {\bibfield  {journal} {\bibinfo  {journal} {Journal of molecular biology}\
  }\textbf {\bibinfo {volume} {288}},\ \bibinfo {pages} {441} (\bibinfo {year}
  {1999})}\BibitemShut {NoStop}%
\bibitem [{\citenamefont {Babin}\ \emph {et~al.}(2008)\citenamefont {Babin},
  \citenamefont {Roland},\ and\ \citenamefont {Sagui}}]{Babin2008}%
  \BibitemOpen
  \bibfield  {author} {\bibinfo {author} {\bibfnamefont {V.}~\bibnamefont
  {Babin}}, \bibinfo {author} {\bibfnamefont {C.}~\bibnamefont {Roland}},\ and\
  \bibinfo {author} {\bibfnamefont {C.}~\bibnamefont {Sagui}},\ }\bibfield
  {title} {\bibinfo {title} {Adaptively biased molecular dynamics for free
  energy calculations},\ }\href@noop {} {\bibfield  {journal} {\bibinfo
  {journal} {The Journal of chemical physics}\ }\textbf {\bibinfo {volume}
  {128}},\ \bibinfo {pages} {134101} (\bibinfo {year} {2008})}\BibitemShut
  {NoStop}%
\bibitem [{\citenamefont {Babin}\ \emph {et~al.}(2009)\citenamefont {Babin},
  \citenamefont {Karpusenka}, \citenamefont {Moradi}, \citenamefont {Roland},\
  and\ \citenamefont {Sagui}}]{Babin2009}%
  \BibitemOpen
  \bibfield  {author} {\bibinfo {author} {\bibfnamefont {V.}~\bibnamefont
  {Babin}}, \bibinfo {author} {\bibfnamefont {V.}~\bibnamefont {Karpusenka}},
  \bibinfo {author} {\bibfnamefont {M.}~\bibnamefont {Moradi}}, \bibinfo
  {author} {\bibfnamefont {C.}~\bibnamefont {Roland}},\ and\ \bibinfo {author}
  {\bibfnamefont {C.}~\bibnamefont {Sagui}},\ }\bibfield  {title} {\bibinfo
  {title} {Adaptively biased molecular dynamics: An umbrella sampling method
  with a time-dependent potential},\ }\href@noop {} {\bibfield  {journal}
  {\bibinfo  {journal} {International Journal of Quantum Chemistry}\ }\textbf
  {\bibinfo {volume} {109}},\ \bibinfo {pages} {3666} (\bibinfo {year}
  {2009})}\BibitemShut {NoStop}%
\bibitem [{\citenamefont {Rosta}\ and\ \citenamefont
  {Hummer}(2015)}]{Rosta2015}%
  \BibitemOpen
  \bibfield  {author} {\bibinfo {author} {\bibfnamefont {E.}~\bibnamefont
  {Rosta}}\ and\ \bibinfo {author} {\bibfnamefont {G.}~\bibnamefont {Hummer}},\
  }\bibfield  {title} {\bibinfo {title} {Free energies from dynamic weighted
  histogram analysis using unbiased markov state model},\ }\href@noop {}
  {\bibfield  {journal} {\bibinfo  {journal} {Journal of chemical theory and
  computation}\ }\textbf {\bibinfo {volume} {11}},\ \bibinfo {pages} {276}
  (\bibinfo {year} {2015})}\BibitemShut {NoStop}%
\bibitem [{\citenamefont {Berendsen}\ \emph {et~al.}(1995)\citenamefont
  {Berendsen}, \citenamefont {van~der Spoel},\ and\ \citenamefont {van
  Drunen}}]{Berendsen1995}%
  \BibitemOpen
  \bibfield  {author} {\bibinfo {author} {\bibfnamefont {H.~J.}\ \bibnamefont
  {Berendsen}}, \bibinfo {author} {\bibfnamefont {D.}~\bibnamefont {van~der
  Spoel}},\ and\ \bibinfo {author} {\bibfnamefont {R.}~\bibnamefont {van
  Drunen}},\ }\bibfield  {title} {\bibinfo {title} {Gromacs: a message-passing
  parallel molecular dynamics implementation},\ }\href@noop {} {\bibfield
  {journal} {\bibinfo  {journal} {Computer physics communications}\ }\textbf
  {\bibinfo {volume} {91}},\ \bibinfo {pages} {43} (\bibinfo {year}
  {1995})}\BibitemShut {NoStop}%
\bibitem [{\citenamefont {Abraham}\ \emph {et~al.}(2015)\citenamefont
  {Abraham}, \citenamefont {Murtola}, \citenamefont {Schulz}, \citenamefont
  {P{\'a}ll}, \citenamefont {Smith}, \citenamefont {Hess},\ and\ \citenamefont
  {Lindahl}}]{Abraham2015a}%
  \BibitemOpen
  \bibfield  {author} {\bibinfo {author} {\bibfnamefont {M.~J.}\ \bibnamefont
  {Abraham}}, \bibinfo {author} {\bibfnamefont {T.}~\bibnamefont {Murtola}},
  \bibinfo {author} {\bibfnamefont {R.}~\bibnamefont {Schulz}}, \bibinfo
  {author} {\bibfnamefont {S.}~\bibnamefont {P{\'a}ll}}, \bibinfo {author}
  {\bibfnamefont {J.~C.}\ \bibnamefont {Smith}}, \bibinfo {author}
  {\bibfnamefont {B.}~\bibnamefont {Hess}},\ and\ \bibinfo {author}
  {\bibfnamefont {E.}~\bibnamefont {Lindahl}},\ }\bibfield  {title} {\bibinfo
  {title} {Gromacs: High performance molecular simulations through multi-level
  parallelism from laptops to supercomputers},\ }\href@noop {} {\bibfield
  {journal} {\bibinfo  {journal} {SoftwareX}\ }\textbf {\bibinfo {volume}
  {1}},\ \bibinfo {pages} {19} (\bibinfo {year} {2015})}\BibitemShut {NoStop}%
\bibitem [{\citenamefont {Tribello}\ \emph {et~al.}(2014)\citenamefont
  {Tribello}, \citenamefont {Bonomi}, \citenamefont {Branduardi}, \citenamefont
  {Camilloni},\ and\ \citenamefont {Bussi}}]{Tribello2014}%
  \BibitemOpen
  \bibfield  {author} {\bibinfo {author} {\bibfnamefont {G.~A.}\ \bibnamefont
  {Tribello}}, \bibinfo {author} {\bibfnamefont {M.}~\bibnamefont {Bonomi}},
  \bibinfo {author} {\bibfnamefont {D.}~\bibnamefont {Branduardi}}, \bibinfo
  {author} {\bibfnamefont {C.}~\bibnamefont {Camilloni}},\ and\ \bibinfo
  {author} {\bibfnamefont {G.}~\bibnamefont {Bussi}},\ }\bibfield  {title}
  {\bibinfo {title} {Plumed 2: New feathers for an old bird},\ }\href@noop {}
  {\bibfield  {journal} {\bibinfo  {journal} {Computer Physics Communications}\
  }\textbf {\bibinfo {volume} {185}},\ \bibinfo {pages} {604} (\bibinfo {year}
  {2014})}\BibitemShut {NoStop}%
\bibitem [{\citenamefont {Lindorff-Larsen}\ \emph {et~al.}(2010)\citenamefont
  {Lindorff-Larsen}, \citenamefont {Piana}, \citenamefont {Palmo},
  \citenamefont {Maragakis}, \citenamefont {Klepeis}, \citenamefont {Dror},\
  and\ \citenamefont {Shaw}}]{Lindorff-Larsen2010}%
  \BibitemOpen
  \bibfield  {author} {\bibinfo {author} {\bibfnamefont {K.}~\bibnamefont
  {Lindorff-Larsen}}, \bibinfo {author} {\bibfnamefont {S.}~\bibnamefont
  {Piana}}, \bibinfo {author} {\bibfnamefont {K.}~\bibnamefont {Palmo}},
  \bibinfo {author} {\bibfnamefont {P.}~\bibnamefont {Maragakis}}, \bibinfo
  {author} {\bibfnamefont {J.~L.}\ \bibnamefont {Klepeis}}, \bibinfo {author}
  {\bibfnamefont {R.~O.}\ \bibnamefont {Dror}},\ and\ \bibinfo {author}
  {\bibfnamefont {D.~E.}\ \bibnamefont {Shaw}},\ }\bibfield  {title} {\bibinfo
  {title} {Improved side-chain torsion potentials for the amber ff99sb protein
  force field},\ }\href@noop {} {\bibfield  {journal} {\bibinfo  {journal}
  {Proteins: Structure, Function, and Bioinformatics}\ }\textbf {\bibinfo
  {volume} {78}},\ \bibinfo {pages} {1950} (\bibinfo {year}
  {2010})}\BibitemShut {NoStop}%
\bibitem [{\citenamefont {Jorgensen}\ \emph {et~al.}(1983)\citenamefont
  {Jorgensen}, \citenamefont {Chandrasekhar}, \citenamefont {Madura},
  \citenamefont {Impey},\ and\ \citenamefont {Klein}}]{Jorgensen1983}%
  \BibitemOpen
  \bibfield  {author} {\bibinfo {author} {\bibfnamefont {W.~L.}\ \bibnamefont
  {Jorgensen}}, \bibinfo {author} {\bibfnamefont {J.}~\bibnamefont
  {Chandrasekhar}}, \bibinfo {author} {\bibfnamefont {J.~D.}\ \bibnamefont
  {Madura}}, \bibinfo {author} {\bibfnamefont {R.~W.}\ \bibnamefont {Impey}},\
  and\ \bibinfo {author} {\bibfnamefont {M.~L.}\ \bibnamefont {Klein}},\
  }\bibfield  {title} {\bibinfo {title} {Comparison of simple potential
  functions for simulating liquid water},\ }\href@noop {} {\bibfield  {journal}
  {\bibinfo  {journal} {The Journal of chemical physics}\ }\textbf {\bibinfo
  {volume} {79}},\ \bibinfo {pages} {926} (\bibinfo {year} {1983})}\BibitemShut
  {NoStop}%
\bibitem [{\citenamefont {Wang}\ \emph {et~al.}(2006)\citenamefont {Wang},
  \citenamefont {Wang}, \citenamefont {Kollman},\ and\ \citenamefont
  {Case}}]{Wang2006}%
  \BibitemOpen
  \bibfield  {author} {\bibinfo {author} {\bibfnamefont {J.}~\bibnamefont
  {Wang}}, \bibinfo {author} {\bibfnamefont {W.}~\bibnamefont {Wang}}, \bibinfo
  {author} {\bibfnamefont {P.~A.}\ \bibnamefont {Kollman}},\ and\ \bibinfo
  {author} {\bibfnamefont {D.~A.}\ \bibnamefont {Case}},\ }\bibfield  {title}
  {\bibinfo {title} {Automatic atom type and bond type perception in molecular
  mechanical calculations},\ }\href@noop {} {\bibfield  {journal} {\bibinfo
  {journal} {Journal of molecular graphics and modelling}\ }\textbf {\bibinfo
  {volume} {25}},\ \bibinfo {pages} {247} (\bibinfo {year} {2006})}\BibitemShut
  {NoStop}%
\bibitem [{\citenamefont {Wang}\ \emph {et~al.}(2004)\citenamefont {Wang},
  \citenamefont {Wolf}, \citenamefont {Caldwell}, \citenamefont {Kollman},\
  and\ \citenamefont {Case}}]{Wang2004a}%
  \BibitemOpen
  \bibfield  {author} {\bibinfo {author} {\bibfnamefont {J.}~\bibnamefont
  {Wang}}, \bibinfo {author} {\bibfnamefont {R.~M.}\ \bibnamefont {Wolf}},
  \bibinfo {author} {\bibfnamefont {J.~W.}\ \bibnamefont {Caldwell}}, \bibinfo
  {author} {\bibfnamefont {P.~A.}\ \bibnamefont {Kollman}},\ and\ \bibinfo
  {author} {\bibfnamefont {D.~A.}\ \bibnamefont {Case}},\ }\bibfield  {title}
  {\bibinfo {title} {Development and testing of a general amber force field},\
  }\href@noop {} {\bibfield  {journal} {\bibinfo  {journal} {Journal of
  computational chemistry}\ }\textbf {\bibinfo {volume} {25}},\ \bibinfo
  {pages} {1157} (\bibinfo {year} {2004})}\BibitemShut {NoStop}%
\bibitem [{\citenamefont {Da~Silva}\ and\ \citenamefont
  {Vranken}(2012)}]{SousadaSilva2012}%
  \BibitemOpen
  \bibfield  {author} {\bibinfo {author} {\bibfnamefont {A.~W.~S.}\
  \bibnamefont {Da~Silva}}\ and\ \bibinfo {author} {\bibfnamefont {W.~F.}\
  \bibnamefont {Vranken}},\ }\bibfield  {title} {\bibinfo {title}
  {Acpype-antechamber python parser interface},\ }\href@noop {} {\bibfield
  {journal} {\bibinfo  {journal} {BMC research notes}\ }\textbf {\bibinfo
  {volume} {5}},\ \bibinfo {pages} {1} (\bibinfo {year} {2012})}\BibitemShut
  {NoStop}%
\bibitem [{\citenamefont {Bussi}\ \emph {et~al.}(2007)\citenamefont {Bussi},
  \citenamefont {Donadio},\ and\ \citenamefont {Parrinello}}]{bussi2007}%
  \BibitemOpen
  \bibfield  {author} {\bibinfo {author} {\bibfnamefont {G.}~\bibnamefont
  {Bussi}}, \bibinfo {author} {\bibfnamefont {D.}~\bibnamefont {Donadio}},\
  and\ \bibinfo {author} {\bibfnamefont {M.}~\bibnamefont {Parrinello}},\
  }\bibfield  {title} {\bibinfo {title} {Canonical sampling through velocity
  rescaling},\ }\href@noop {} {\bibfield  {journal} {\bibinfo  {journal} {J.
  Chem. Phys.}\ }\textbf {\bibinfo {volume} {126}},\ \bibinfo {pages} {014101}
  (\bibinfo {year} {2007})}\BibitemShut {NoStop}%
\bibitem [{\citenamefont {Parrinello}\ and\ \citenamefont
  {Rahman}(1981)}]{parrinello1981}%
  \BibitemOpen
  \bibfield  {author} {\bibinfo {author} {\bibfnamefont {M.}~\bibnamefont
  {Parrinello}}\ and\ \bibinfo {author} {\bibfnamefont {A.}~\bibnamefont
  {Rahman}},\ }\bibfield  {title} {\bibinfo {title} {Polymorphic transitions in
  single crystals: A new molecular dynamics method},\ }\href@noop {} {\bibfield
   {journal} {\bibinfo  {journal} {Journal of Applied physics}\ }\textbf
  {\bibinfo {volume} {52}},\ \bibinfo {pages} {7182} (\bibinfo {year}
  {1981})}\BibitemShut {NoStop}%
\end{thebibliography}%
